\documentclass[sigconf]{acmart}

\AtBeginDocument{%
  \providecommand\BibTeX{{%
    \normalfont B\kern-0.5em{\scshape i\kern-0.25em b}\kern-0.8em\TeX}}}

\usepackage{hyperref}
\usepackage[skip=8pt]{caption}
\usepackage{subcaption}
\usepackage{graphicx}
\usepackage{array}
\usepackage{multicol}
\usepackage{multirow}
\usepackage{longtable}
\usepackage{makecell}
\usepackage{enumitem}
\setlist{nolistsep}
\usepackage{color}
\usepackage{xspace}
\usepackage{rotating}
\usepackage{amsmath}

\newcommand{\lblb}{OPT$_{lb}\_$EVAL$_{lb}$\xspace}
\newcommand{\lbmb}{OPT$_{lb}\_$EVAL$_{mb}$\xspace}
\newcommand{\mbmb}{OPT$_{mb}\_$EVAL$_{mb}$\xspace}

\copyrightyear{2024}
\acmYear{2024}
\setcopyright{rightsretained}
\acmConference[E-Energy '24]{The 15th ACM International Conference on Future and Sustainable Energy Systems}{June 4--7, 2024}{Singapore, Singapore}
\acmBooktitle{The 15th ACM International Conference on Future and Sustainable Energy Systems (E-Energy '24), June 4--7, 2024, Singapore, Singapore}\acmDOI{10.1145/3632775.3639587}
\acmISBN{979-8-4007-0480-2/24/06}

\begin{document}

\title[The Green Mirage]{The Green Mirage: Impact of Location- and Market-based Carbon Intensity Estimation on Carbon Optimization Efficacy}

\author{Diptyaroop Maji}
\affiliation{%
  \institution{University of Massachusetts Amherst}
  \city{}
  \country{}}

\author{Noman Bashir}
\affiliation{%
  \institution{Massachusetts Institute of Technology}
  \city{}
  \country{}}

  \author{David Irwin}
\affiliation{%
  \institution{University of Massachusetts Amherst}
  \city{}
  \country{}}

\author{Prashant Shenoy}
\affiliation{%
  \institution{University of Massachusetts Amherst}
  \city{}
  \country{}}

  \author{Ramesh K. Sitaraman}
\affiliation{%
  \institution{University of Massachusetts Amherst}
  \city{}
  \country{}}

\renewcommand{\shortauthors}{Maji et al.}

\begin{abstract}
  In recent years, there has been an increased emphasis on reducing the carbon emissions from electricity consumption. 
Many organizations have set ambitious targets to reduce the carbon footprint of their operations as a part of their sustainability goals. 
The carbon footprint of any consumer of electricity is computed as the product of the total energy consumption and the carbon intensity of electricity. 
Third-party carbon information services provide information on carbon intensity across regions that consumers can leverage to modulate their energy consumption patterns to reduce their overall carbon footprint. 
In addition, to accelerate their decarbonization process, large electricity consumers increasingly acquire power purchase agreements (PPAs) from renewable power plants to obtain renewable energy credits that offset their ``brown'' energy consumption.

There are primarily two methods for attributing carbon-free energy, or renewable energy credits, to electricity consumers: location-based and market-based. These two methods yield significantly different carbon intensity values for various consumers. As there is a lack of consensus which method to use for carbon-free attribution, a concurrent application of both approaches is observed in practice.
In this paper, we show that such concurrent applications can cause discrepancies in the carbon savings reported by carbon optimization techniques. Our analysis across three state-of-the-art carbon optimization techniques shows \textcolor{black}{possible overestimation of up to 55.1\%} in the carbon reductions reported by the consumers and even increased emissions for consumers in some cases. We also find that carbon optimization techniques make different decisions under the market-based method and location-based method, and \textcolor{black}{the market-based method can yield up to $28.2 \%$ less carbon savings than those claimed by the location-based method for consumers without PPAs.}

\end{abstract}



\maketitle

\section{Introduction}
\label{sec:introduction}
\begin{figure}[t]
    \centering
    \includegraphics[width=0.8\linewidth]{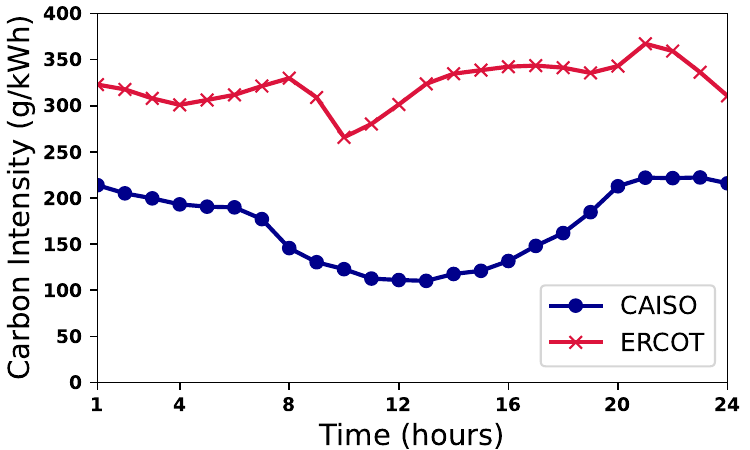}
    \vspace{-0.1cm}
    \caption{\emph{Carbon intensity of electricity varies spatially and temporally, with regions and periods with more renewable energy having lower carbon intensity values.}}
    \vspace{-0.4cm}
    \label{fig:ciso_vs_ercot_ci}
\end{figure}

Recently, there has been an increased emphasis on decarbonizing the electric grid, that has resulted in significant increases in the deployment of renewable energy sources, such as hydro, geothermal, solar, and wind energy. The deployment of low-carbon electricity sources reduces the overall carbon intensity (CI) of electricity --- measured as grams of ${CO}_2eq$ emitted per $kWh$ of electricity generated or consumed --- which in turn reduces the overall emissions from electricity consumption. Until recently, the electric grid was opaque and did not expose the mix of generation sources or the carbon intensity of supplied electricity to its consumers. However, the emergence of third-party services such as Electricity Maps~\cite{emap} and WattTime~\cite{watttime} in recent years has enabled consumers to receive real-time carbon intensity information. As shown in Figure~\ref{fig:ciso_vs_ercot_ci}, these services can provide real-time estimates and forecasts of carbon intensity, which indicate how green the electricity supply is in any location at any specified time.

Importantly, businesses and researchers are increasingly using these carbon intensity estimates and forecasts to modulate their electricity usage and reduce their carbon emissions. Such approaches seek to exploit spatial and temporal variations in the carbon intensity, which occur due to differences in the generation source mix (and different penetration of renewables) across regions and changes caused by the grid dispatch schedule at each location. One popular carbon optimization technique is based on time shifting, which moves flexible loads from high to low carbon intensity periods. Examples include deferring electric vehicle (EV) charging or delaying the execution of batch jobs to low-carbon periods of the day. Another class of techniques has focused on spatial shifting, where computing workloads like machine learning training are moved to cloud regions with the greenest electricity. Since grid electricity is unlikely to be entirely carbon-free for several decades, such approaches are gaining in popularity for reducing the carbon footprint of an organization or an individual.

A key prerequisite for the above carbon optimization approaches is the availability of \emph{accurate} carbon intensity information. The Greenhouse Gas (GHG) Protocol recognizes two primary methods for attributing carbon-free energy~\cite{scope2-ghg-guidance}: the location-based method and the market-based method. Each method yields a different carbon intensity metric, which we denote as $CI_{lb}$ and $CI_{mb}$ for the location-based and market-based methods, respectively. \emph{Location-based} (LB) attribution assumes that the electricity consumed by each consumer is based on all  generation sources (both fossil-based and fossil-free) present at that location. In other words, this method assumes that it is impossible to segregate the exact source that generated the electricity consumed by each consumer, and all consumers are assumed to use the same proportion of energy mix as the grid's generation mix.

In contrast, \emph{market-based} (MB) attribution assumes a consumer or organization can choose the generation source that supplies their electricity, even though it is technically infeasible to route specific electricity generation to specific end-consumers even if both are on the same grid.  Hence, the attribution is done via market-based accounting mechanisms, where an organization can claim renewable energy exclusively for their needs by purchasing Renewable Energy Credits (RECs)~\cite{epa_rec} or through Power Purchase Agreements (PPAs)~\cite{ppa, financial_ppa}. Consequently, consumers who have signed PPAs will see a carbon intensity based on their renewable energy purchases as well as any additional grid usage, while the remaining consumers see a different carbon intensity based on the remainder of the generation sources (residual mix) at that location.

Carbon reduction algorithms can be classified based on which attribution method they use for their optimization and which method they use for evaluating their carbon reduction. \textcolor{black}{We use the terms OPT$_{lb}$ and OPT$_{mb}$ to denote location-based and market-based optimization and assume that they use $CI_{lb}$ and $CI_{mb}$ carbon intensity metrics, respectively, for their optimization. Likewise, we use the terms EVAL$_{lb}$ (resp., EVAL$_{mb}$) for algorithms that use $CI_{lb}$ (resp, $CI_{mb}$) for evaluating their carbon reduction.  Much of the current research on spatial and temporal shifting has optimized and evaluated its benefits under the location-based method (i.e., scenario \lblb).} Such approaches have not considered the growing use of PPAs in the energy markets and have not evaluated the carbon emission reduction under the market-based method. Further, the GHG protocol states that all parties should report their emissions using both location- and market-based methods, but they can use either of the methods to evaluate their carbon reduction to meet their carbon emission  goals~\cite{scope2-ghg-guidance, brander2018creative}. Thus, we see instances of both methods used by different parties, leading to discrepancies. For example, carbon intensity services such as Electricity Maps~\cite{emap} and WattTime~\cite{watttime} provide location-based carbon intensity ($CI_{lb}$). At the same time, some of the renewable sources at that location may be contracted through PPAs and organizations with those PPAs prefer to use the market-based method, resulting in the same green sources being attributed to multiple consumers. \textcolor{black}{To prevent this, we should use the market-based attribution ($CI_{mb}$) to evaluate carbon reductions of consumers without PPAs, although these consumers optimize based on location-based carbon intensity ($CI_{lb}$). Thus, their reported carbon reduction based on location-based evaluation (i.e., \lblb) could differ significantly from their actual carbon reduction after accounting for the PPAs (i.e., scenario \lbmb).}

As a result, the two metrics for carbon intensity based on the two attribution methods can yield different results. However, accurate attribution and carbon intensity estimation are critical for doing carbon-aware optimization and properly understanding the benefits under different methods.

\noindent
{\bf Our Contributions.} In this paper, we highlight the differences and challenges in using the different carbon intensity metrics and the resulting impact on popular carbon optimization techniques. We take a data-driven approach where we first consider the carbon reduction benefits under the location-based method (\lblb). We evaluate the discrepancies that can arise as carbon-free energy is claimed by a subset of consumers using PPAs (\lbmb). Finally, we quantify the carbon reduction benefits if such optimization approaches account for PPAs and optimize and evaluate using the market-based method (i.e., scenario \mbmb). Specifically, our paper makes the following contributions:
\begin{enumerate}[leftmargin=*, topsep=-0.00cm]
    \item First, we show how the carbon intensity estimation varies depending on the attribution method. Our analysis across 123 regions worldwide shows that the estimates can vary by up to $194\%$ if all solar and wind energy in a region is contracted out.

    \item \textcolor{black}{Next, we look at techniques optimizing for carbon using the location-based method and analyze the discrepancy in carbon reduction between location- and market-based evaluations (\lblb versus \lbmb). Our analysis of three such techniques shows that location-based evaluations can possibly overestimate carbon reduction by up to $55.1\%$. In some cases, location-based evaluations may even report reductions when there is an increase in carbon emissions for consumers without PPAs under market-based evaluations.}

    \item \textcolor{black}{Finally, we analyze how much carbon reduction these techniques can get when both optimized and evaluated using the market-based method (\mbmb). Our evaluation finds that for consumers without PPAs, \mbmb can result in up to $28.2 \%$ less carbon savings than claimed by prior work under the location-based method (\lblb) if all solar and wind energy is under PPA. We also find that the optimizers make different decisions when using the market-based method than using the location-based method, as $CI_{lb}$ can significantly differ from $CI_{mb}$.}

    
\end{enumerate}

\section{Background}
\label{sec:background}
In this section, we provide background on the carbon intensity of electricity, different carbon attribution methods, and the current state-of-the-art carbon-optimization techniques in use.

\vspace{-0.2cm}
\subsection{Carbon Intensity of Electricity}
\label{sec:carbon-intensity}
Electricity is generated using a mix of conventional non-renewable sources like coal, petroleum, and gas and renewable sources like solar, wind, and hydro. The mix of sources varies across grids of different regions. For example, California depends mainly on solar and natural gas, while hydro is Sweden's main electricity generation source. Electricity demand is time-varying within a region, and the grid must ensure that the supply always matches the demand. Renewable sources like solar and wind are volatile; their generation depends on weather conditions and cannot be controlled. As a result, the grid maintains a set of generators that can be turned on or off quickly to meet the demand, and hence, the source mix also varies temporally. While electricity is also often exchanged between grids, we ignore electricity exchange in this paper for simplicity.

Carbon emitted due to electricity generation is measured in terms of carbon intensity. The average carbon intensity (CI) of electricity is the amount of carbon emitted (in grams) per unit of electrical energy produced or consumed (in kWh). In this paper, we use the terms electricity (or energy) production and electricity (or energy) generation interchangeably. 
The carbon intensity of electricity can be mathematically formulated as a weighted average (refer ~\cite{maji2022carboncast}):
\begin{equation}
    \label{eq:carbonIntensity}
    Average\ Carbon\ Intensity\ (CI_{avg})   = \frac{\sum{(E_i * CEF_i)}}{\sum{E_i}}
\end{equation}

where $E_i$ is the electrical energy produced ($MWh$) by a source $i$ \& $CEF_i$ is the carbon emission factor ($g/kWh$) of that source.


Since the source mix is variable, the carbon intensity of electricity also varies across regions and with time. Figure~\ref{fig:ciso_vs_ercot_ci} shows how the average carbon intensity in California (CAISO) differs from Texas (ERCOT), and how it varies within a day in both regions. Non-renewable sources have higher CEFs than renewable sources. For example, coal and natural gas have CEFs of 760 g/kWh and 370 g/kWh, respectively, while all renewable sources have zero CEF. Since CAISO has a higher renewable penetration than ERCOT, CAISO has a lower average carbon intensity. Also, within CAISO itself, the carbon intensity is lower during the day when there is solar generation than at night when natural gas compensates for the lack of solar and meets the demand.

\vspace{-0.2cm}
\subsection{Carbon Attribution Methods} 
\label{sec:accounting-methods}
\begin{figure}[t]
    \centering
    \includegraphics[width=0.8\linewidth]{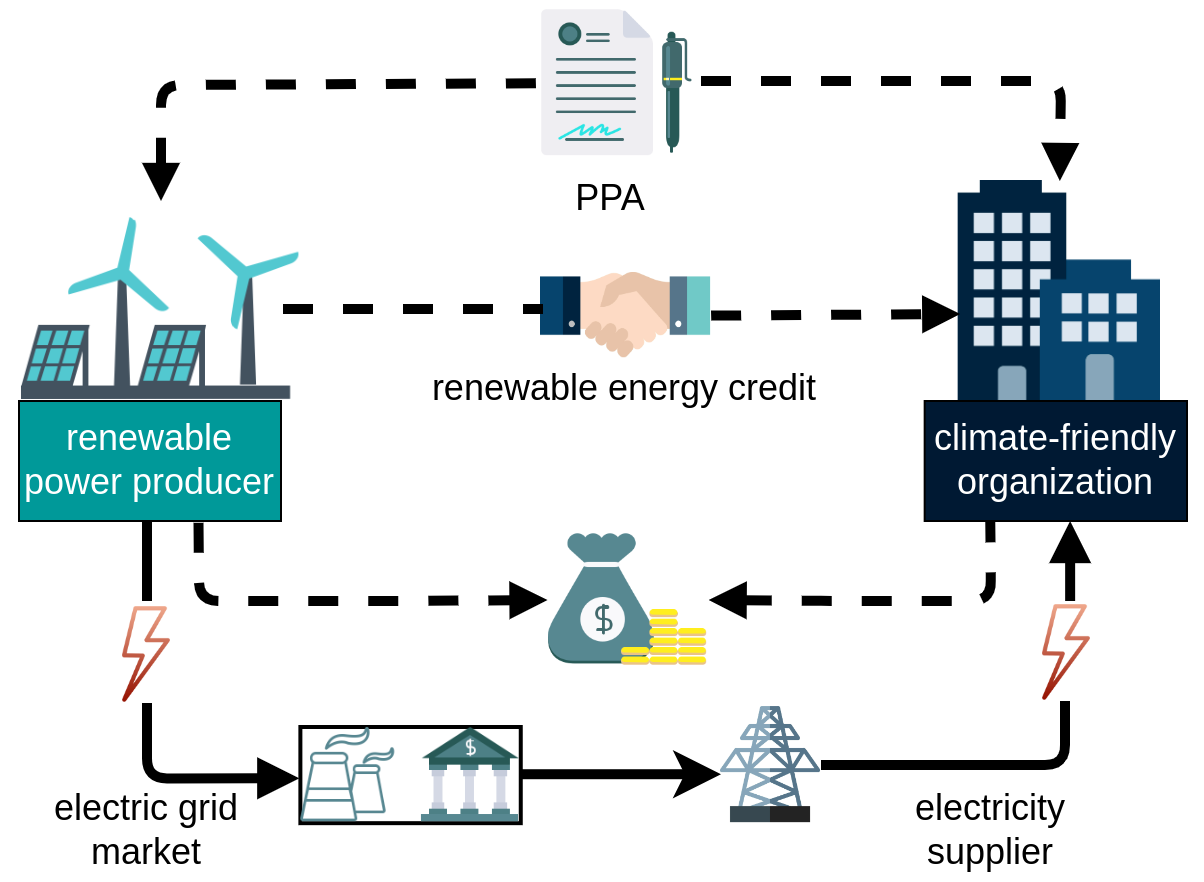}
    \vspace{-0.2cm}
    \caption{\emph{An illustration of the Power Purchase Agreement (PPA) commonly used to meet carbon-free energy targets. \textcolor{black}{Solid lines denote the physical flow of electricity; dashed lines denote the transactional flows.}}}
    \vspace{-0.4cm}
    \label{fig:ppa}
\end{figure}

To mitigate the effect of GHG emissions from electricity, companies are increasingly setting targets to become carbon-neutral~\cite{carbon_neutral} or even net-zero~\cite{un_net_zero}. The term ``net-zero'' refers to achieving an overall balance between GHG emissions produced and emissions taken out of the atmosphere. In this paper, we only consider scope 2 emissions, that is, indirect greenhouse gas (GHG) emissions resulting from the purchase of electricity~\cite{scope2-ghg-guidance}. Few companies are shifting their demand to low-carbon regions or periods to reduce their emissions, but many companies are
investing in renewable energy via Power Purchase Agreements (PPAs). PPAs are usually long-term contracts for renewable energy between a consumer and an electricity producer, wherein the consumer can claim renewable energy credits for their investment (refer Figure~\ref{fig:ppa}) 
and lower the emissions caused by the electricity they consume.
Based on scope 2 GHG guidance protocol~\cite{scope2-ghg-guidance}, there are two methods of attributing the carbon emissions to consumers depending on how the green energy and the source mix of electricity are attributed.

\noindent
\textbf{Location-based (LB) method.} In the location-based method, all the consumers inside a geographical location get the same electricity mix. Green energy is attributed to the grid, and the average carbon intensity of the grid mix, including both renewable and non-renewable sources in proportion to the electricity generated by them, is used. This is done regardless of any green energy investments made by any specific consumer, and all consumers  share any renewable investment made by a particular consumer.

\noindent
\textbf{Market-based (MB) method.} In the market-based method, a consumer who has invested in renewables can claim the credit for the purchased electricity and reduce their carbon emissions even if their invested green energy is not physically delivered via transmission lines. All invested green energy is first attributed to the investing parties, and then the carbon intensity of the residual grid mix without the invested renewables is estimated. Investing consumers meet any remaining demand using the residual mix. On the other hand, consumers without any renewable investments account for their carbon emissions only using the carbon intensity of the residual mix. Thus, a consumer investing in electricity that matches the amount of their energy consumption can claim to be $100 \%$ renewable even if the electricity they physically use comes from a grid mix with renewable and non-renewable sources.



\subsection{Carbon Optimization Techniques}
\label{sec:carbon-optimization-techniques}
Countries, organizations, and even individuals are taking measures to reduce their carbon emissions in recent years. Popular carbon optimization techniques in practice today try to take advantage of the spatial and temporal variations in carbon intensity and do more work when and where the carbon intensity is low. Broadly, there are three categories of optimization techniques in the literature:

\vspace{0.1cm}
\begin{enumerate}[leftmargin=*]
\item \textbf{Spatial load shifting:} Cloud services have geographically distributed data centers (DCs) to provide clients with high availability and fault tolerance. Some carbon optimization techniques have leveraged this fact to redirect client requests to the greenest available data center and thus reduce the carbon emission of computing~\cite{gao2012s, maji2023bringing}. For example, a request serviced in California will emit less carbon than one in Texas (refer Figure~\ref{fig:ciso_vs_ercot_ci}) for the same amount of compute.

\item \textbf{Temporal load shifting:} Cloud providers and applications leverage workload flexibility and schedule workloads at low-carbon periods to reduce the carbon footprint~\cite{radovanovic2022carbon, wiesner2021let}. For example, given a 24-hour deadline for a 2-hour job, scheduling the job during the day in California will emit less carbon than during the night for the same amount of compute (refer Figure~\ref{fig:ciso_vs_ercot_ci}). Beyond the computing industry, there has been work on charging electric vehicles (EVs) during greener times~\cite{cheng2022carbon} to reduce the carbon footprint.

\item \textbf{Resource autoscaling:} For cloud applications with limited temporal and spatial flexibility, the carbon footprint can also be reduced by increasing the resources allocated to an application during low-carbon periods while scaling back the resources during other periods with high carbon intensity~\cite{hanafy2023carbonscaler}. Since the application execution does not need to be delayed, it often finishes faster than the temporal load-shifting technique. 
\end{enumerate}


\section{Carbon Intensity Estimation}
\label{sec:ci-estimation}
In this section, we show how carbon intensity can be estimated in two ways, based on the different carbon-free energy attribution methods discussed in Section~\ref{sec:accounting-methods}, and the effects of PPAs on carbon intensity calculation. We also discuss how these two estimates differ across regions and times. We consider all renewables to be carbon-free, with zero emissions.

\vspace{-0.2cm}
\subsection{Location-Based Carbon Intensity}
When we consider all the sources generating electricity in a region, we call it the total grid mix, and we refer to the associated carbon intensity as the total grid carbon intensity. The location-based attribution method always considers the total grid mix regardless of any renewable investments; hence, the carbon intensity of electricity in this method ($CI_{lb}$) is the same as the total grid carbon intensity and can be calculated using Equation~\ref{eq:carbonIntensity}. That is, 
\begin{equation} \label{eq:total_ci}
    CI_{lb} = \frac{\sum{(E_i \times CEF_i)}}{E}
\end{equation}
where $E$ is the total electricity production in a grid, and $\sum(E_i) = E$. The value of the carbon intensity is \emph{same for everyone who consumes electricity from that grid}.

\subsection{Market-Based Carbon Intensity}
\label{sec:ci_res}
If some renewable (carbon-free) sources are contracted out via PPAs, we refer to the remaining source mix in the grid as the residual grid mix, and we call the associated carbon intensity residual carbon intensity ($CI_{res}$). In this paper, we only consider \emph{solar and wind energy} when we say renewables under contract since these are the primary sources used in PPAs today. \textcolor{black}{In a grid, suppose $E_i$ be the total electricity generated (MWh) by a source $i$, out of which let $E^{ppa}_i$ be the electricity contracted through PPA. Suppose $E$ be the total electricity generated in that grid, and $E_{ppa}$ be the total electricity under PPA. Therefore, $\sum E_i = E$, and $\sum E^{ppa}_i = E_{ppa}$. In the market-based method, entities with PPAs claim the greenness of their invested electricity. Thus, all the contracted carbon-free energy is removed before calculating $CI_{res}$. Then, 
\begin{equation} \label{eq:residual_ci}
    CI_{res} = \frac{\sum{((E_i-E^{ppa}_i) \times CEF_i)}}{E-E_{ppa}},
\end{equation}
Therefore, the carbon intensity of a consumer using the market-based method ($CI_{mb}$) in that grid can be obtained using
\begin{equation} \label{eq:mb_ci}
    CI_{mb} = (1-f)\times CI_{res}
\end{equation}
where $f\in[0, 1]$ is the fraction of electricity consumption met by PPA. Thus, contrary to the location-based method, the \emph{market-based carbon intensity varies across the consumers in the same grid}. For consumers with no PPAs ($f=0$), $CI_{mb} = CI_{res}$, whereas for a consumer able to meet their total demand via PPAs, $CI_{mb} = 0$. The market-based method always favors the investors of carbon-free energy over non-investors.}

\begin{figure}[t]
    \begin{center}
       \includegraphics[width=0.8\linewidth]{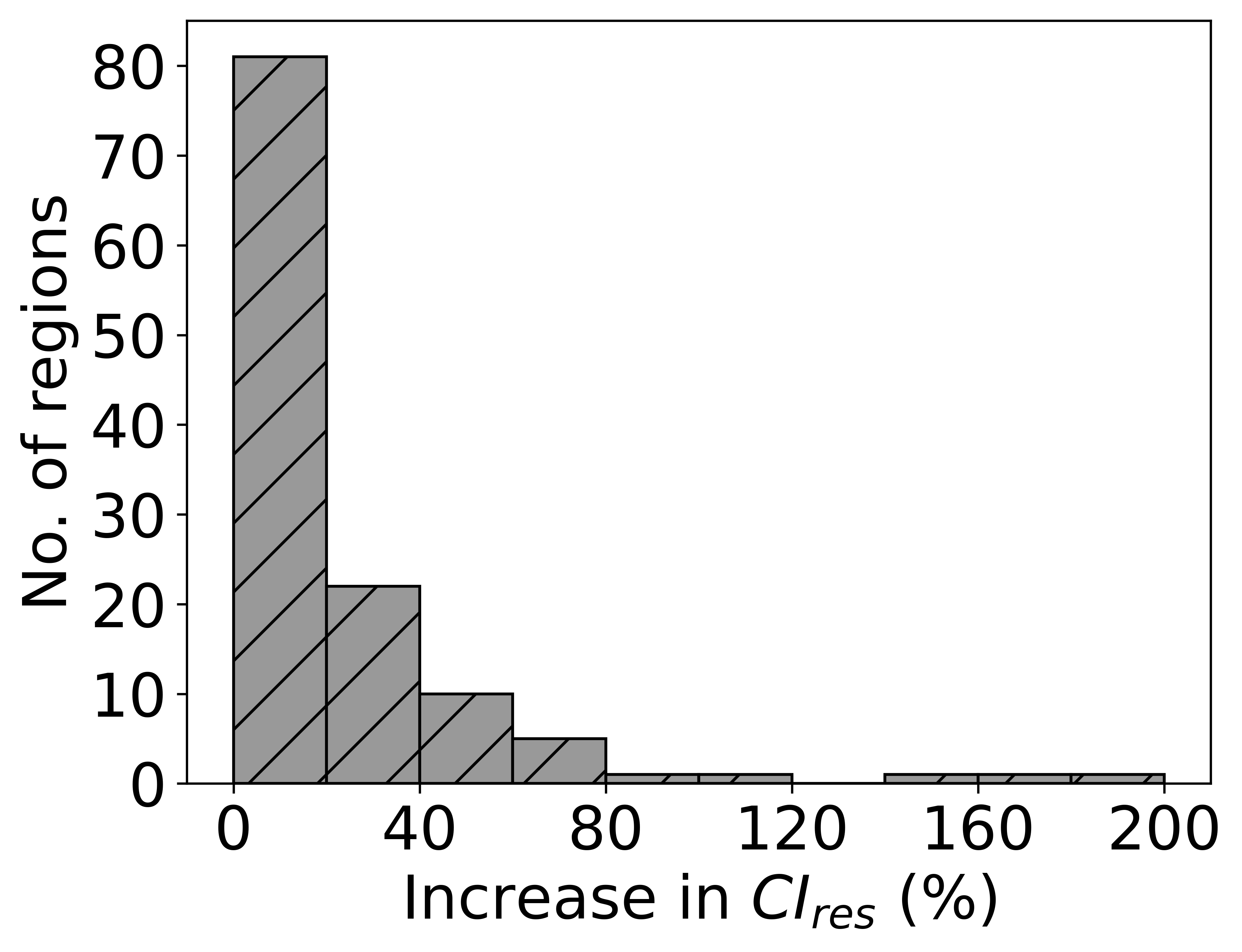}
    \end{center}
    \vspace{-0.3cm}
       \caption{\emph{Average increase in $CI_{res}$ across 123 regions worldwide when all solar and wind energy is contracted out.}}
    \label{fig:residual_vs_solar_wind}
    \vspace{-0.3cm}
\end{figure}

\label{sec:res_vs_tot_ci}
\begin{figure}[t]
    \begin{center}
       \includegraphics[width=0.9\linewidth]{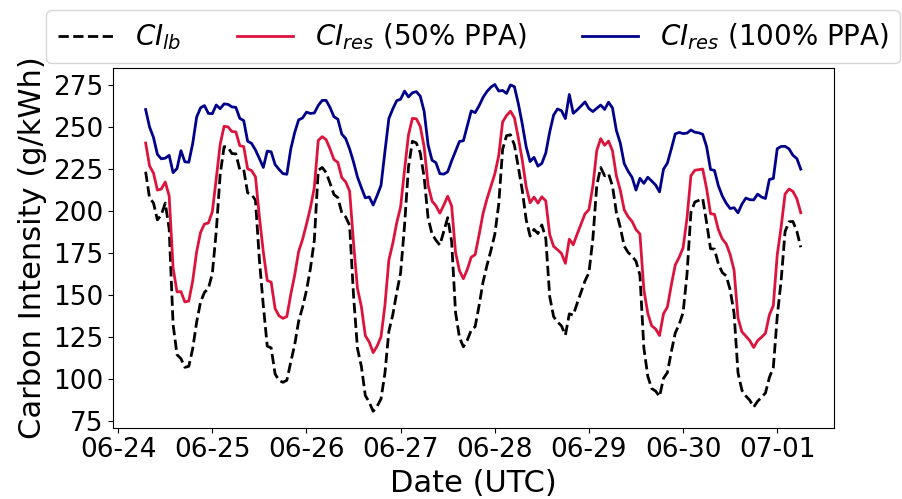}
    \end{center}
    \vspace{-0.3cm}
       \caption{\emph{Weekly trace showing how $CI_{res}$ in California differs from $CI_{lb}$ as more and more renewables are contracted out.}}
    \label{fig:residual_CI_increase_week}
    \vspace{-0.5cm}
\end{figure}

\subsection{Differences between Location-Based and Market-Based Carbon Intensity}

In this paper, we only consider consumers without PPAs for our analysis. Since $CI_{mb} = CI_{res}$ for those consumers, we use these notations interchangeably. $CI_{res}$ will naturally be higher than $CI_{lb}$ since the fraction of renewables in the residual mix will be smaller than in the total mix. We now show how $CI_{res}$ differs from $CI_{lb}$ across regions and even temporally within a region. For this analysis, we used grid and carbon intensity data from 123 regions worldwide (data obtained from Electricity Maps~\cite{emap}).

Figure~\ref{fig:residual_vs_solar_wind} shows a histogram of the percentage increase in $CI_{res}$ over $CI_{lb}$ for all the regions when all renewables are contracted out. The regions with high solar and wind penetration show a high increase in $CI_{res}$. The median increase across all the considered regions is $11.9\%$, with the highest difference being $194\%$ in South Australia. As solar and wind penetration increases in most grids worldwide along with the amount of electricity purchased via PPAs, this difference will only increase with time.

Within a region, $CI_{res}$ increases with increasing amounts of purchased renewable energy. We show how the carbon intensity changes in the California grid (CAISO) over a week as more and more renewables are purchased (refer Figure~\ref{fig:residual_CI_increase_week}). CAISO grid generates a large fraction of electricity from solar, and hence, its carbon intensity is much lower in the day when solar energy is available. However, with an increasing PPA percentage, not only does $CI_{res}$ increase, but temporal variations in carbon intensity also decrease.

\begin{figure}[t]
    \begin{center}
       \includegraphics[width=0.9\linewidth]{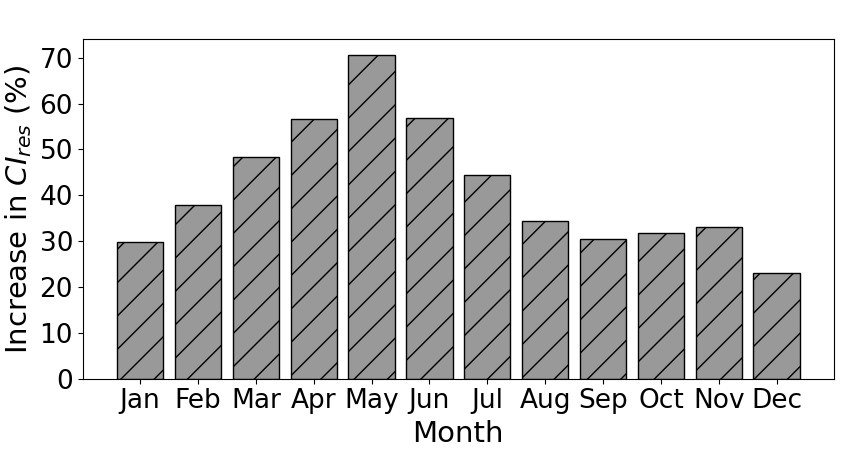}
    \end{center}
    \vspace{-0.3cm}
       \caption{\emph{In California, the difference between $CI_{res}$ and $CI_{lb}$ varies seasonally, with spring and summer months showing more difference when there is more solar energy.}}
    \label{fig:residual_CI_increase_month}
    \vspace{-0.5cm}
\end{figure}

Figure~\ref{fig:residual_CI_increase_month} shows the difference between $CI_{res}$ and $CI_{lb}$ across the different months of 2022, assuming all solar and wind energy is purchased. Since CAISO has a lot of solar-generated electricity, the difference is far greater during spring and summer than during winter. This is because, during spring and summer, a greater fraction of electricity is generated from solar as there is more sunlight. We observe a similar effect diurnally, with the difference being higher during the day than at night (refer Figure~\ref{fig:ciso_daily_variation}).

\begin{figure}[t]
    \begin{center}
       \includegraphics[width=0.8\linewidth]{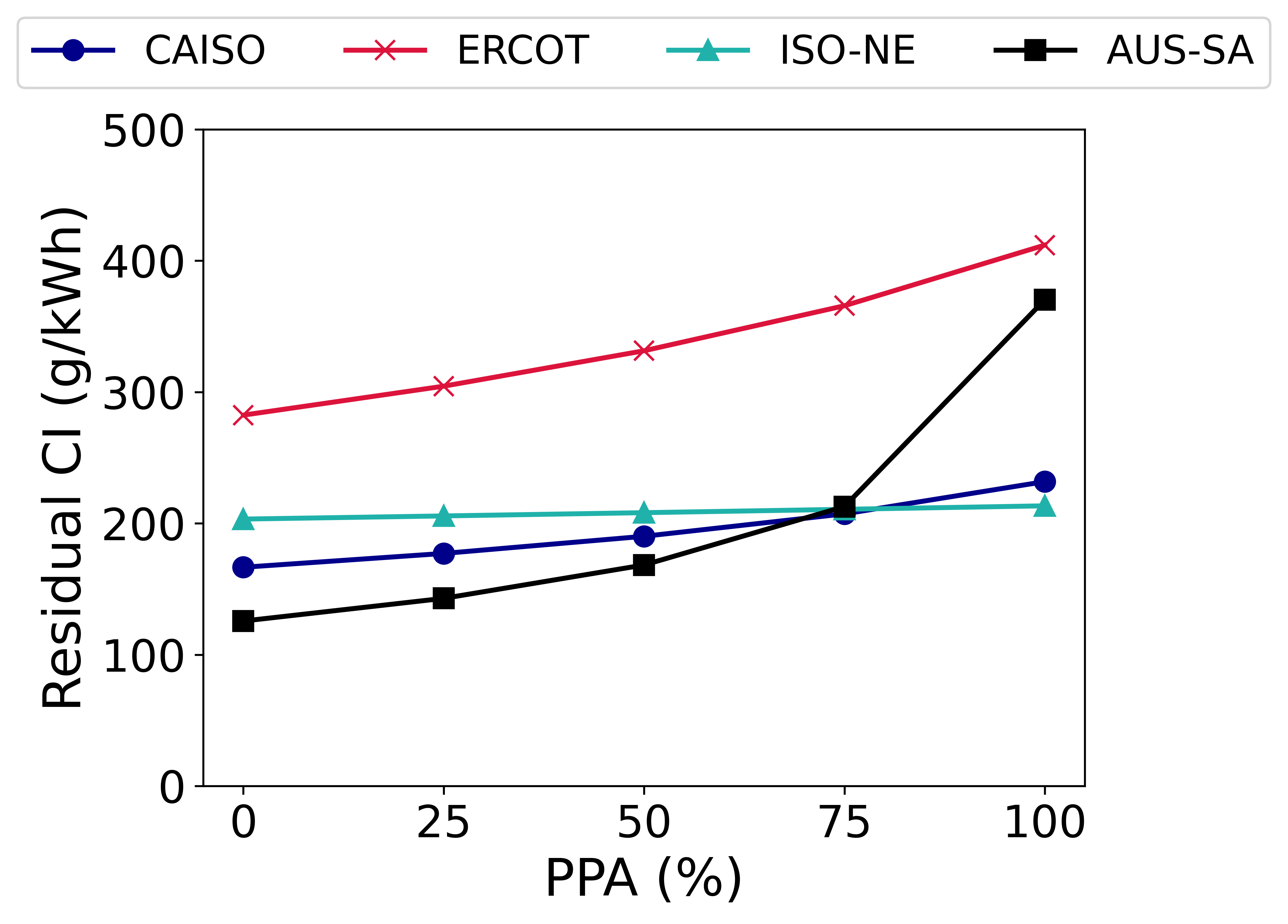}
    \end{center}
    \vspace{-0.3cm}
       \caption{\emph{Carbon intensity of the residual electricity increases with the fraction of PPA-contracted solar and wind capacity.}}
    \label{fig:residual_CI_increase}
    \vspace{-0.5cm}
\end{figure}

As $CI_{res}$ increases with increasing amounts of contracted renewable energy, sometimes a region having lower $CI_{lb}$ (or lower $CI_{res}$ initially) than another region may have a higher $CI_{res}$ than the second region. For example, in Figure~\ref{fig:residual_CI_increase}, both in South Australia (AUS-SA) and in California's CAISO grid, $CI_{res}$ becomes higher than that of New England's ISO-NE grid beyond a certain point as more and more renewables are contracted out. This is because AUS-SA and CAISO have a much higher proportion of solar and wind, which makes the residual energy browner faster than the ISO-NE grid.

Thus, $CI_{res}$ varies significantly from $CI_{lb}$ across the regions and seasonally (or diurnally) even within a region, with some low-carbon regions becoming browner than other high-carbon regions as more and more carbon-free energy is contracted out.

\subsection{Carbon Intensity Estimation in Practice}
\label{ci_practice}

The two attribution methods provide two carbon intensity metrics that differ significantly. For example, Electricity Maps' data \cite{emap_no} shows that the electricity mix in Norway comprises 99\% renewables. Hence, $CI_{lb}$ in Norway using location-based attribution is nearly zero. However, renewable generators in Norway sell electricity to consumers in other countries (who are in physically disjoint grids), and the residual mix is only left with 15\% renewable sources~\cite{aib_residual}, resulting in a much higher $CI_{res}$. Thus, even though consumers in Norway consume nearly carbon-free electricity, they can not claim to  consume that green energy under the market-based attribution rules (since the renewable credits have been sold to other parties).

If everyone followed either location- or market-based carbon attribution, there will be no discrepancy~\cite{holzapfel2023electricity}. However, in practice, there is no consensus on which method to use, resulting in some organizations using one approach while others using the alternate one. The concurrent use of both approaches by different entities can result in
discrepancies in the carbon emission reductions reported by the consumers. The buyers of PPAs can claim the resulting carbon-free energy for themselves to meet a portion of their electricity demand and rely on third-party carbon intensity services~\cite{emap, watttime, maji2022dacf, maji2022carboncast} to estimate the carbon emissions of their remaining demand. Consumers without PPAs rely on these services to estimate the carbon emissions of their entire demand. However, current carbon information services typically may not have information about the residual mix and provide location-based carbon intensity $CI_{lb}$ estimates. This happens mainly due to the lack of availability of real-time data --- data about the residual grid mix often lags by years~\cite{aib_residual, green_e_residual} --- but this is relatively common in the present scenario~\cite{google_247_cfe}. Consequently, renewables under PPAs in a region may be counted doubly by different or sometimes even the same set of consumers. As shown in Section~\ref{sec:res_vs_tot_ci}, since $CI_{res}$ varies significantly from $CI_{lb}$, this can result in an overestimation of the ``greenness'' of electric grids, leading to discrepancies in carbon emissions reported by the consumers. 

The concurrent use of both attribution methods due to a lack of visibility into $CI_{res}$ also presents significant challenges when performing carbon-aware optimizations. If a consumer without PPAs uses $CI_{lb}$ for optimization instead of $CI_{res}$ when others have already claimed the contracted electricity, it can lead to sub-optimal decision-making and often overestimated amount of reported carbon emissions reductions. In the following section, we explain the effects of different carbon intensity signals on current state-of-the-art carbon optimizations in the computing industry.

\section{Discrepancies in Carbon-savings with concurrent attribution}
\label{sec:implications-ppa-unaware}
As mentioned in Section~\ref{sec:carbon-optimization-techniques}, there has been significant research in the computing industry and beyond to reduce the carbon footprint by leveraging the spatial and temporal variations of carbon intensity~\cite{radovanovic2022carbon, wiesner2021let, gao2012s, maji2023bringing, hanafy2023carbonscaler, cheng2022carbon}. All these techniques depend on accurate knowledge of real-time and forecasted carbon intensity. Section~\ref{sec:ci-estimation} highlighted how the current carbon intensity estimation approaches can lead to discrepancies in the face of location- and market-based mechanisms. Carbon optimization techniques use third-party carbon intensity services~\cite{emap, watttime, maji2022dacf, maji2022carboncast} for decision-making, which use the location-based method and are inherently PPA-unaware. However, the reported savings only apply if everyone follows the location-based method or if no electricity is contracted. In reality, a fraction of carbon-free energy is under PPA in any region and claimed by electricity buyers. So, that energy should not be counted doubly when others are estimating their carbon emissions. 


\begin{figure*}[ht]
    \begin{subfigure}{0.33\linewidth}
     \begin{center}
     \includegraphics[width=\textwidth]{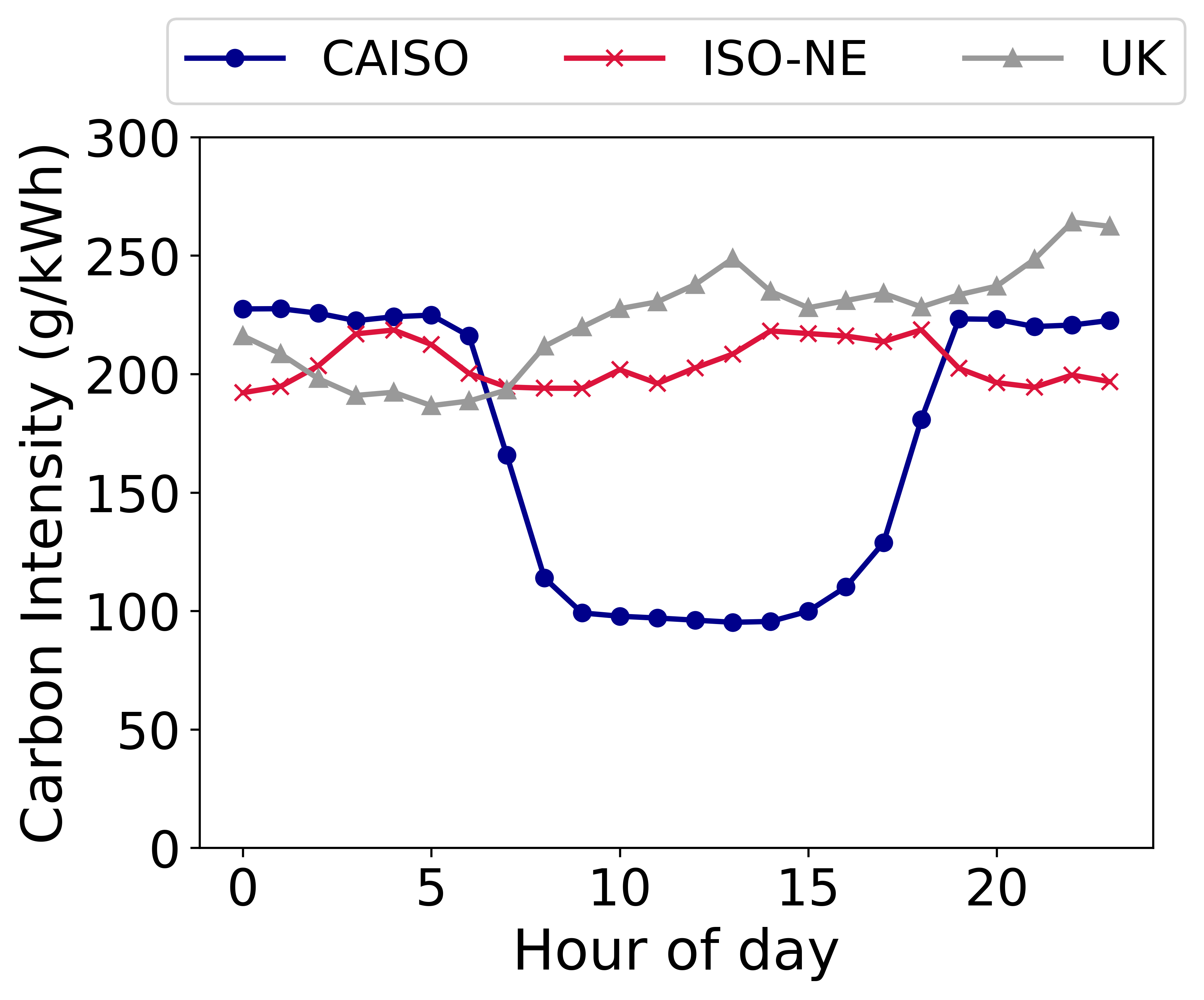}
     \end{center}
     \caption{Total carbon intensity \\($CI_{lb}$).}
     \label{fig:spatial_grid_ci}
    \end{subfigure}
    \begin{subfigure}{0.33\linewidth}
     \begin{center}
     \includegraphics[width=\textwidth]{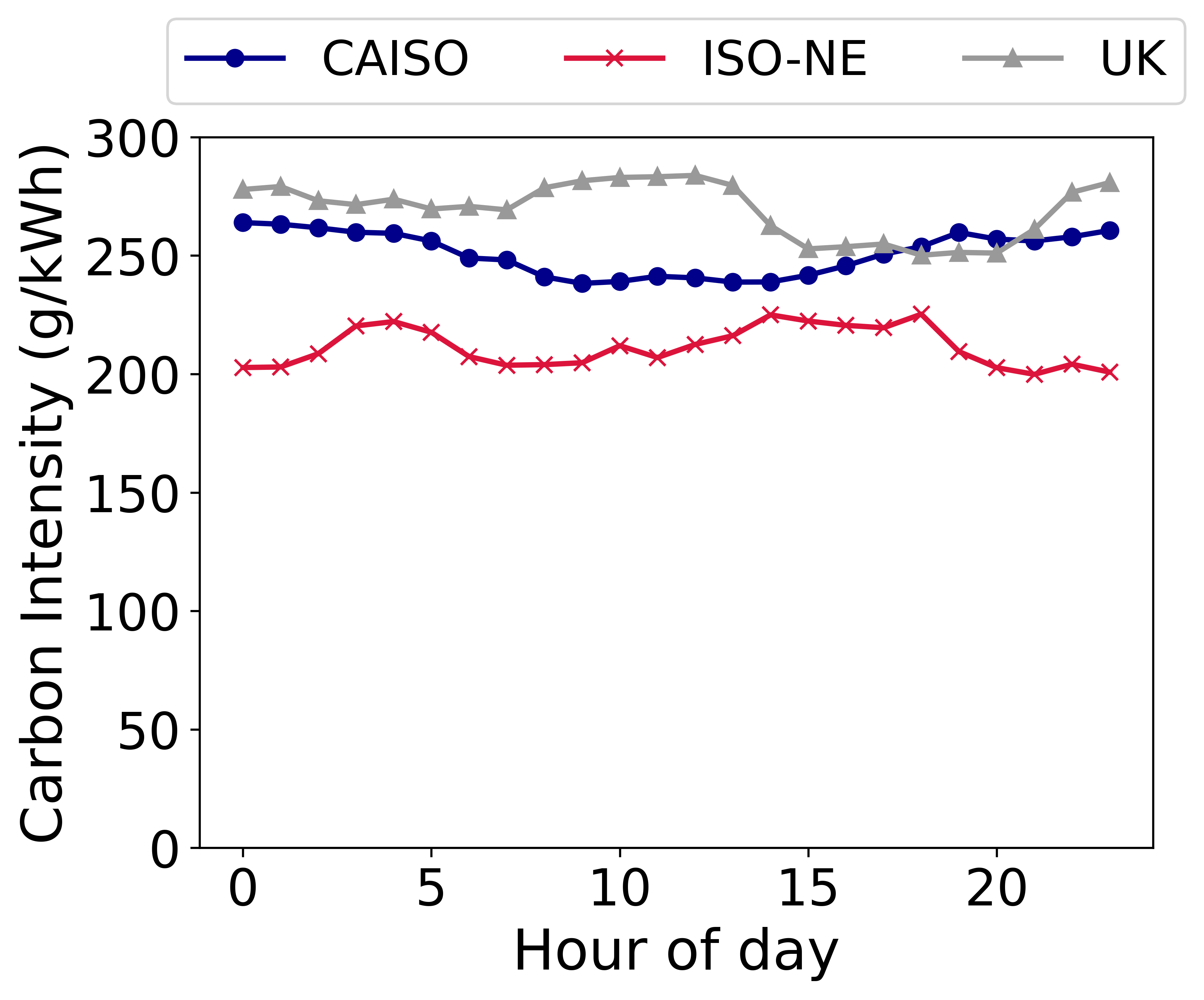}
     \end{center}
     \caption{Residual carbon intensity \\($CI_{res}$).}
     \label{fig:spatial_residual_ci}
    \end{subfigure}
    \begin{subfigure}{0.33\linewidth}
     \begin{center}
     \includegraphics[width=\textwidth]{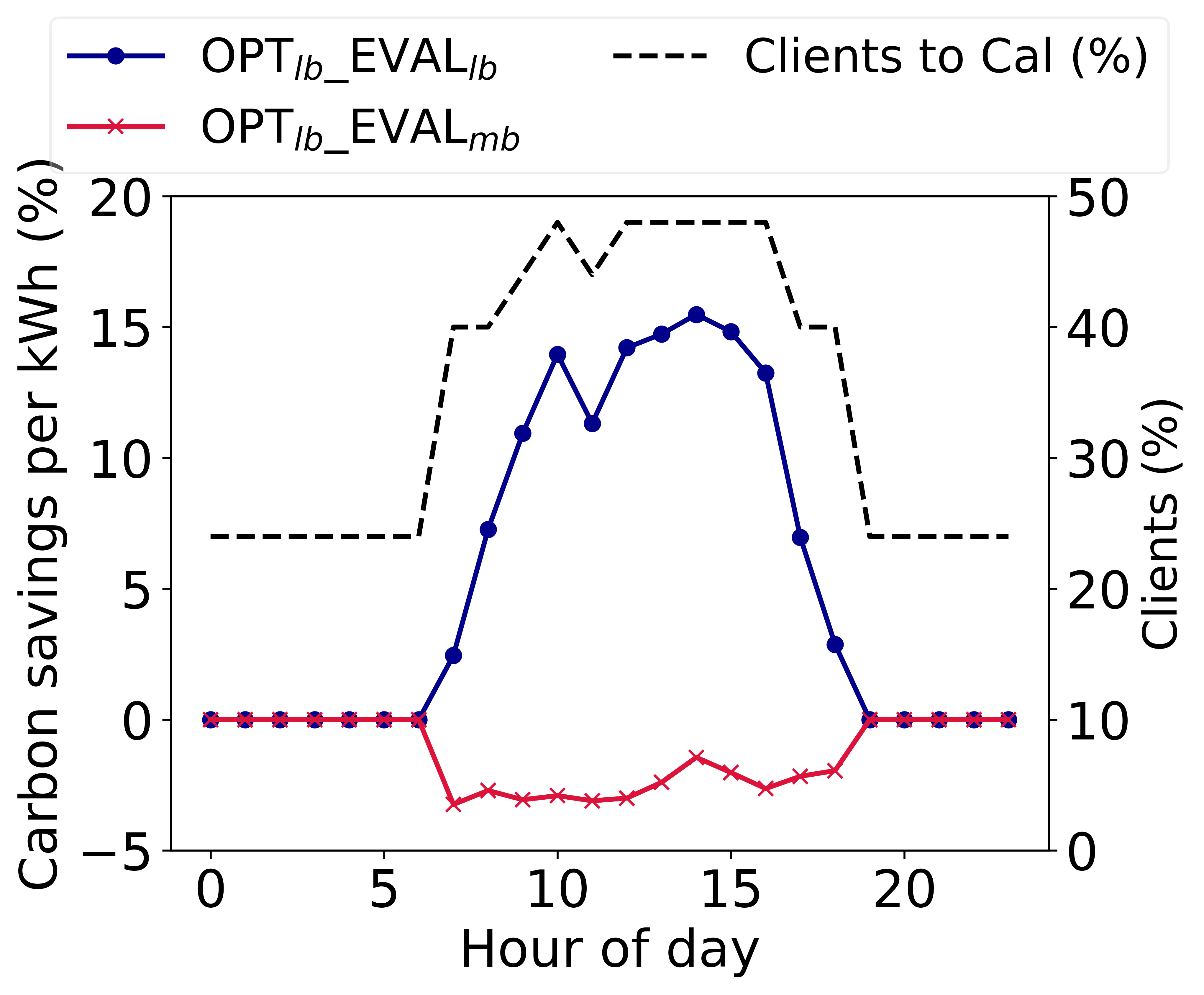}
     \end{center}
     \caption{Carbon savings range and percentage of clients redirected to California.}
     \label{fig:spatial_best_worst}
    \end{subfigure}
     \caption{If all renewables are under PPA, the $CI_{res}$ of CAISO becomes higher than ISO-NE's. Thus, shifting loads to California while not considering PPAs can incur up to $17.2\%$ discrepancy in carbon savings reported by consumers without PPAs.}
     \label{fig:spatial_grid_vs_residual}
     \vspace{-0.3cm}
\end{figure*}


\textcolor{black}{In this section, we show the discrepancy between the carbon savings reported when a carbon optimization technique makes decisions and evaluates based on $CI_{lb}$ (\lblb) and the carbon savings obtained with the same decisions but evaluated with $CI_{mb}$ (\lbmb). We only show this from the viewpoint of consumers without any PPAs. \textcolor{black}{We hypothesize that the discrepancies will be similar for consumers whose PPAs are insufficient to meet their demand but to a lesser extent. However, detailed analysis for consumers with PPAs is kept as future work.}} We show this for the following three state-of-the-art decarbonization techniques:
\begin{itemize}
    \item Carbon-aware spatial load shifting.
    \item Carbon-aware temporal load shifting.
    \item Carbon-aware resource autoscaling.
\end{itemize}

\textcolor{black}{When evaluated using the market-based method, carbon savings decrease with an increasing PPA percentage and can even be negative in some cases. That is, PPA-unaware decisions made using these techniques can increase carbon emissions for the consumer. Since PPA information and $CI_{res}$ are not accessible in real time, this is what is potentially happening in practice. In this section, we show the extreme case where all renewables are contracted out, thus resulting in the maximum possible discrepancy between location- and market-based evaluations.}

\subsection{Spatial Load Shifting}
\label{sec:lb-spatial}
Recently, Maji et al.~\cite{maji2023bringing} have developed a prototype for a carbon-aware load balancer which redirects client requests to greener regions while maintaining latency constraints. Specifically, given N geographically distributed data centers, they use the following equation to determine the optimal server to server a client request:
\begin{equation}\label{score_func}
    DC = min(0.67\times{CI}_i + 0.33\times d_i),\  i= 1\ldots N,
\end{equation}
where $CI_i$ is the carbon intensity in the region where $DC_i$ resides, and $d_i$ is client-to-$DC_i$ distance. They used marginal carbon intensity data from WattTime~\cite{watttime} for their optimization and for calculating carbon emissions and reported an average reduction of $21\%$ carbon emissions (\lblb) over a carbon-unaware load balancer that redirects clients only to the closest data center.

We simulated their technique and compared the carbon savings against a carbon-unaware load balancer. Since their code and data are proprietary, we developed a representative implementation of their algorithm. We also used average carbon intensity data from Electricity Maps~\cite{emap} instead of marginal carbon intensity. We believe this is a valid replacement as their optimization works with both marginal and average carbon intensities, and both Electricity Maps and WattTime~\cite{watttime} estimate carbon intensity using the location-based method~\cite{emap_data}. We considered three data centers in California (CAISO), New England (ISO-NE) and the UK and distributed clients across the US and Europe. We then simulated for a day in 2022.

\textcolor{black}{Figure~\ref{fig:spatial_best_worst} shows the results. The carbon-aware load balancer reduces carbon emissions by up to $15.4\% $ per kWh of electricity consumed over a carbon-unaware load balancer when optimizing and evaluating using the location-based method (\lblb). It achieves this by redirecting more clients to California (CAISO) during the day when solar is available, and $CI_{lb}$ of CAISO is much lower than other regions (refer Figure~\ref{fig:spatial_grid_ci}). However, not considering the renewables under PPA while deciding where to redirect clients will raise discrepancies. If all renewables in the data center regions are under PPA, $CI_{res}$ of CAISO is higher than that of ISO-NE (refer Figure~\ref{fig:spatial_residual_ci}). In that case, redirecting more clients to California reduces carbon savings. We see that using market-based evaluations, the carbon-aware load balancer may potentially emit up to $3.1\%$ more carbon per kWh than a carbon-unaware load balancer over the course of the day. The maximum discrepancy we observe on that day is $17.2\%$ per kWh of electricity consumed when, according to location-based evaluation (\lblb), there are $14.2\%$ carbon savings, but according to market-based evaluation (\lbmb), there is a $3\%$ carbon emission increase. Thus, a consumer without any PPAs would have the impression of reducing emissions, but due to concurrent attribution and a lack of visibility into $CI_{res}$, carbon emissions have potentially increased.}

Note that the discrepancy is per kWh of electricity consumed. Hence, the increased carbon emissions in grams will potentially be significant for large-scale data centres consuming electricity in MWh. Also, we showed the analysis for a day and for the above-mentioned data center regions. Depending on the time of the year and the location of the data centers, this discrepancy can possibly be even higher. For example, in South Australia, $CI_{res}$ can be up to $194\%$ more than $CI_{lb}$ (refer Section~\ref{sec:res_vs_tot_ci}). So, data centers in South Australia will likely observe a much higher discrepancy.

\subsection{Temporal Load Shifting}
\label{sec:lb-temporal}
Next, we show the discrepancy in a carbon-aware job scheduling algorithm that executes flexible workloads during greener periods to reduce the carbon footprint. Wiesner et al.~\cite{wiesner2021let} use carbon intensity forecasts to gain advanced knowledge of low-carbon periods and execute periodically scheduled jobs during those periods. They reported that for nightly jobs scheduled at 1 a.m. daily, with a flexibility window of $\pm 8$ hours, carbon emissions could be reduced by $34.8\%$ in California by scheduling the job during low-carbon hours within that $\pm 8$-hour window. Their carbon intensity forecasts are estimated using the location-based method, which does not consider PPAs (\lblb). Hence, there is a discrepancy in the carbon savings if some percentage of electricity is under PPA. We see that when PPAs are accounted for during evaluation (\lbmb), and if all renewable energy is under PPA, using the same flexibility window and job schedule as above only achieves a carbon saving of $10.3\%$ in California --- a $24.5\%$ discrepancy.

\begin{figure}[t]
    \begin{center}
       \includegraphics[width=0.8\linewidth]{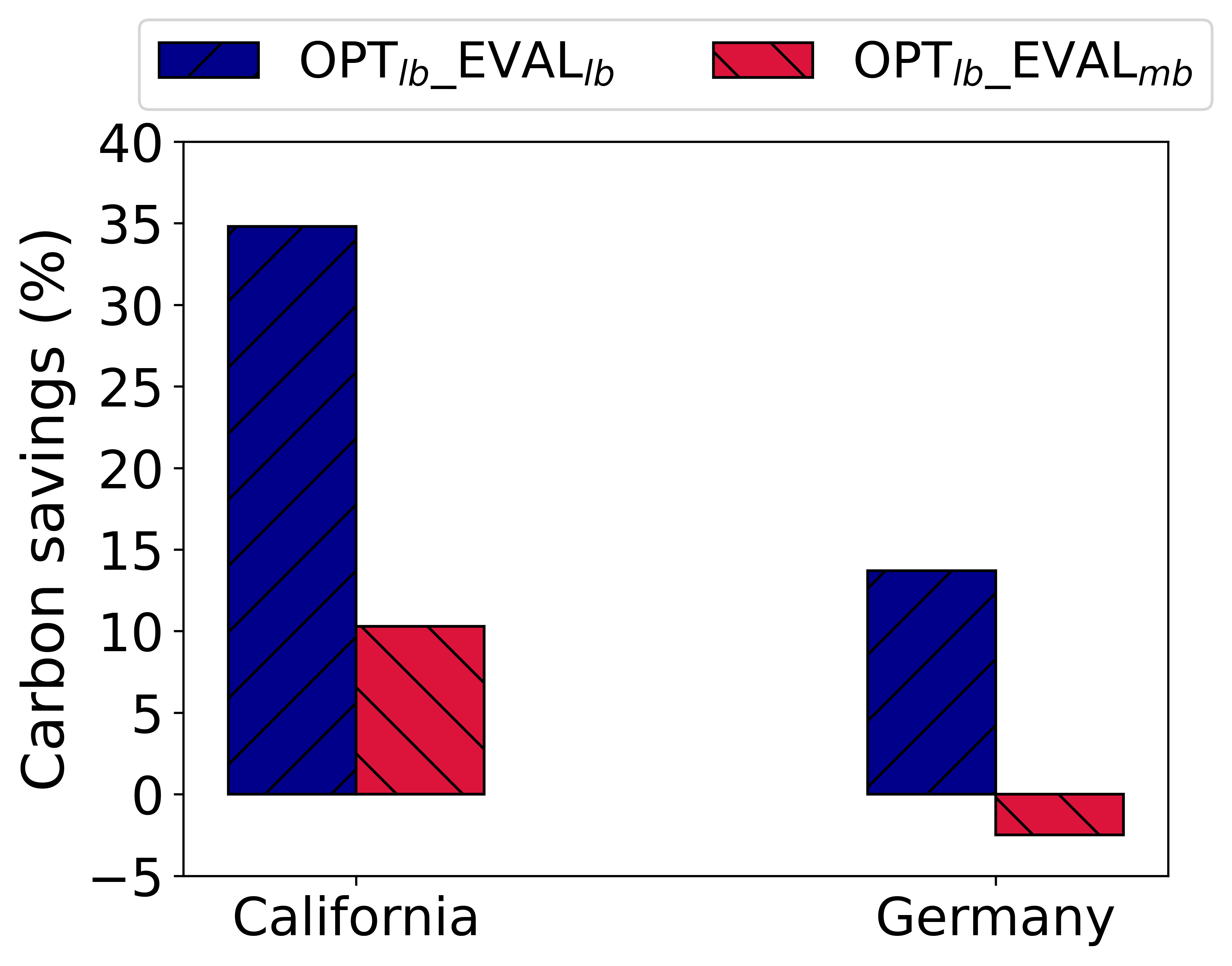}
    \end{center}
    \vspace{-0.2cm}
       \caption{\emph{Scheduling jobs without considering PPAs raises discrepancies in carbon savings. In Germany, this leads to $2.9\%$ more emissions compared to a carbon-unaware job scheduler.}}
    \label{fig:temporal_best_worst}
    \vspace{-0.3cm}
\end{figure}

\textcolor{black}{In Germany, location-based optimization and evaluation reduce emissions by $13.7 \%$ for the same job and flexibility window. However, considering PPAs during evaluation (\lbmb) results in up to $2.5\%$ emission increase for the same schedule. Although the $16.2\%$ discrepancy observed here is less than that in California, the situation in Germany is worse as it potentially results in more emissions than a carbon-unaware baseline due to this discrepancy. Figure~\ref{fig:temporal_best_worst} shows the results for California and Germany.}

\begin{figure}[t]
    \begin{center}
       \includegraphics[width=0.8\linewidth]{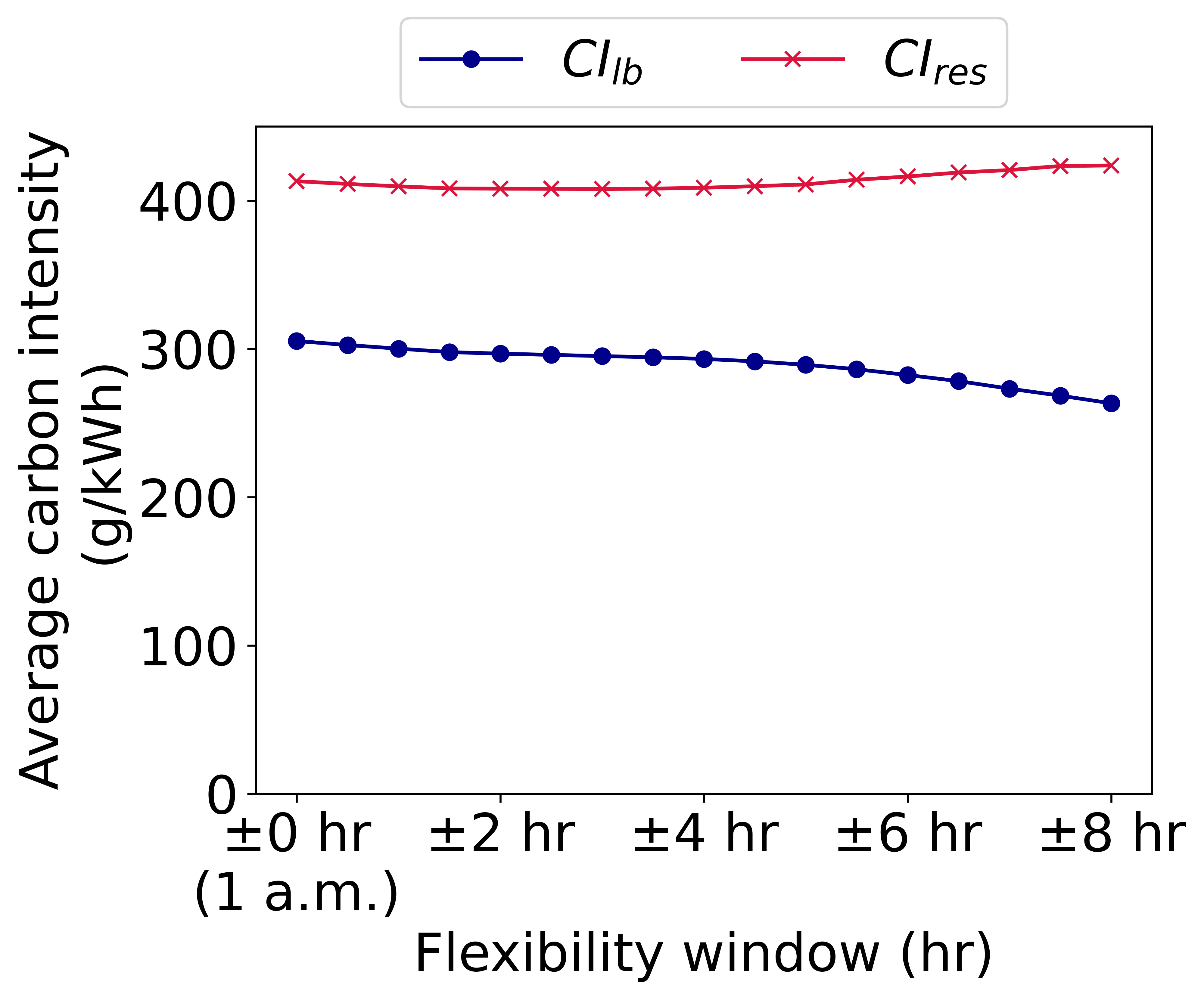}
    \end{center}
    \vspace{-0.2cm}
       \caption{\emph{$CI_{lb}$ and $CI_{res}$ show opposite trends as the flexibility window increases, causing increased emissions in Germany when PPAs are not considered.}}
    \label{fig:temporal_ger_actual_vs_reported}
    \vspace{-0.3cm}
\end{figure}

We see the reason for this flip in Germany in Figure~\ref{fig:temporal_ger_actual_vs_reported}. $CI_{lb}$ gets lower as the flexibility window increases to $\pm 8$ hours, while the average $CI_{res}$ at the hours within the flexibility window when the job is scheduled is higher than the $CI_{res}$ at 1 a.m. Thus, for temporal load shifting optimizations, if $CI_{lb}$ and $CI_{res}$ follow opposite trends, a consumer without PPAs would report reduced emissions while, in reality, the emissions may increase.

\begin{figure}[t]
    \begin{center}
       \includegraphics[width=0.75\linewidth]{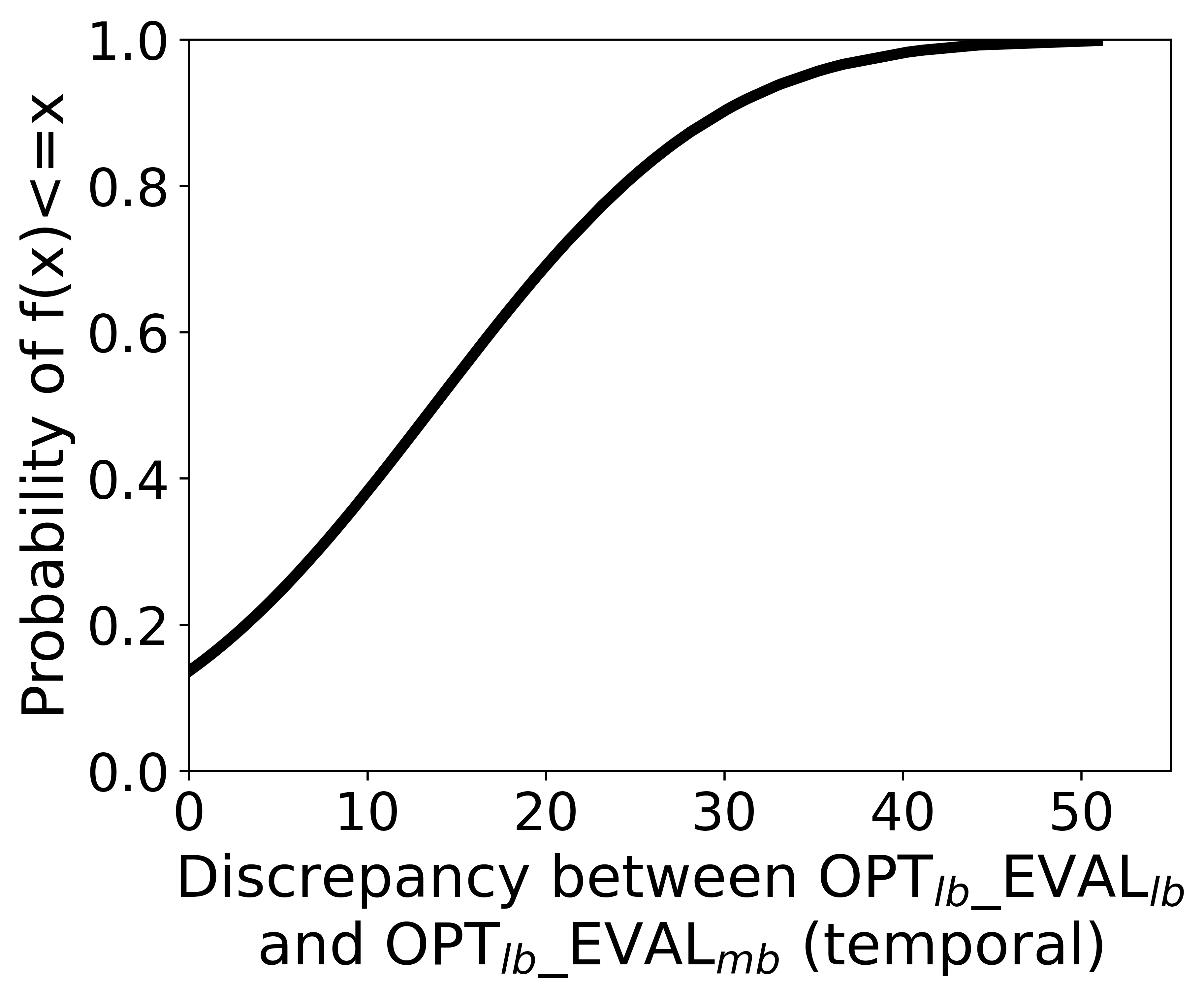}
    \end{center}
    \vspace{-0.2cm}
       \caption{\emph{CDF plot showing that the discrepancies due to carbon-aware temporal shifting can be up to $50.8\%$ when analyzed over Electricity Maps'~\cite{emap} 2022 data.}}
    \label{fig:temporal-discrepancy-cdf}
    \vspace{-0.3cm}
\end{figure}

\textcolor{black}{Since we have grid data from 123 regions, we analyze the discrepancy between location- and market-based evaluations in more detail by applying Wiesner et al.'s~\cite{wiesner2021let} location-based optimizing algorithm on over 100 regions using Electricity Maps'~\cite{emap} data for 2022. Some regions are excluded either due to data inconsistencies or because those regions are powered entirely by renewables. We keep the job start times fixed at 1 a.m. but increase the flexibility window to $\pm 12$ hours so that jobs can be scheduled any time within 24 hours. Figure~\ref{fig:temporal-discrepancy-cdf} is a CDF plot of the discrepancies, showing an average discrepancy of $13.7 \%$ and a maximum potential discrepancy of $50.8 \%$. 
Generally, regions with higher differences between $CI_{res}$ and $CI_{lb}$ show more discrepancies. Note that some regions do not show any discrepancy because there is no electricity generated from solar or wind in those regions (hence, $CI_{res} = CI_{lb}$).}

\subsection{Resource Autoscaling}
\label{sec:lb-autoscale}

\begin{figure}[t]
    \begin{center}
       \includegraphics[width=0.9\linewidth]{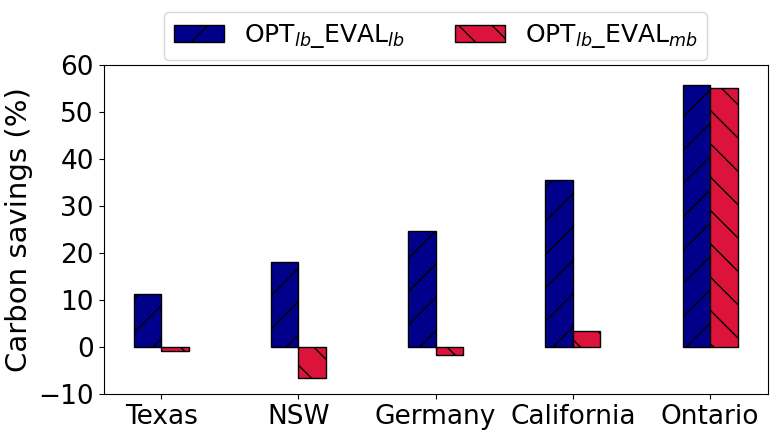}
    \end{center}
    \vspace{-0.2cm}
       \caption{\emph{Discrepancies in carbon savings arising due to PPA-unaware autoscaling, shown for five cloud regions.}}
    \label{fig:scale_savings}
    \vspace{-0.3cm}
\end{figure}

Hanafy et al. have developed CarbonScaler~\cite{hanafy2023carbonscaler}, a system that allocates more resources to cloud applications during low-carbon periods to reduce their carbon footprint. They use carbon intensity forecasts obtained from Electricity Maps~\cite{emap}. Since PPAs are not considered, they follow location-based optimization and show similar discrepancies as in Sections~\ref{sec:lb-spatial} and ~\ref{sec:lb-temporal}. To show that discrepancy, we obtained the CarbonScaler code and simulated a 24-hour ML job (training ResNet18~\cite{he2016deep} model). The job is interruptible and flexible but has to be completed within 24 hours. We simulated this experiment for five \emph{AWS cloud regions} around the world. For each cloud region, we simulated starting at different hours throughout 2022 and reported the average carbon savings over a baseline that always uses one resource instance. In our experiments, we set the maximum resource instances available to CarbonScaler to eight. That is, during low-carbon periods, at most eight instances can run simultaneously to speed up the job execution.

\textcolor{black}{Figure~\ref{fig:scale_savings} shows the results. When evaluating using location-based method (\lblb), we obtained an average savings of $29.2\%$ across the five regions over the baseline that do not scale up resources. However, assuming all electricity is under PPA, evaluating using the market-based method (\lbmb) shows that the same scaling schedule obtains a reduced savings of $9.9\%$ across those regions. The average discrepancy across the regions is $19.2\%$, with a maximum potential discrepancy of $32.2\%$ in California. Additionally, in Texas, New South Wales (NSW), and Germany, carbon emissions may possibly increase if CarbonScaler allocates resources without considering PPAs. Note that the average discrepancy would have been higher if we did not include Ontario, which has very little solar and wind energy and hence is unaffected even though PPAs are not considered.}

\begin{figure}[t]
    \begin{center}
       \includegraphics[width=0.85\linewidth]{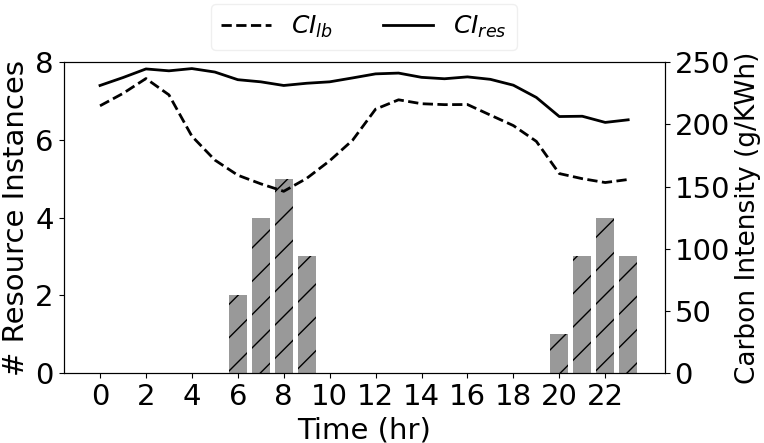}
    \end{center}
    \vspace{-0.2cm}
       \caption{\emph{CarbonScaler~\cite{hanafy2023carbonscaler} following $CI_{lb}$ allocates more instances during high $CI_{res}$, resulting in reduced carbon savings and sometimes increased carbon emissions.}}
    \label{fig:scale_task_schedule}
    \vspace{-0.3cm}
\end{figure}

\textcolor{black}{We now focus on California to explain the reason behind such overestimation (Figure~\ref{fig:scale_task_schedule}). We take a particular day where location-based evaluation reports carbon savings, but market-based evaluation shows increased emissions compared to the baseline. On this day, following $CI_{lb}$ and not considering PPAs, CarbonScaler allocates more resources when the $CI_{lb}$ is lower. However, $CI_{res}$ is much higher at those hours, resulting in increased emissions.}

\begin{figure}[t]
    \begin{center}
       \includegraphics[width=0.75\linewidth]{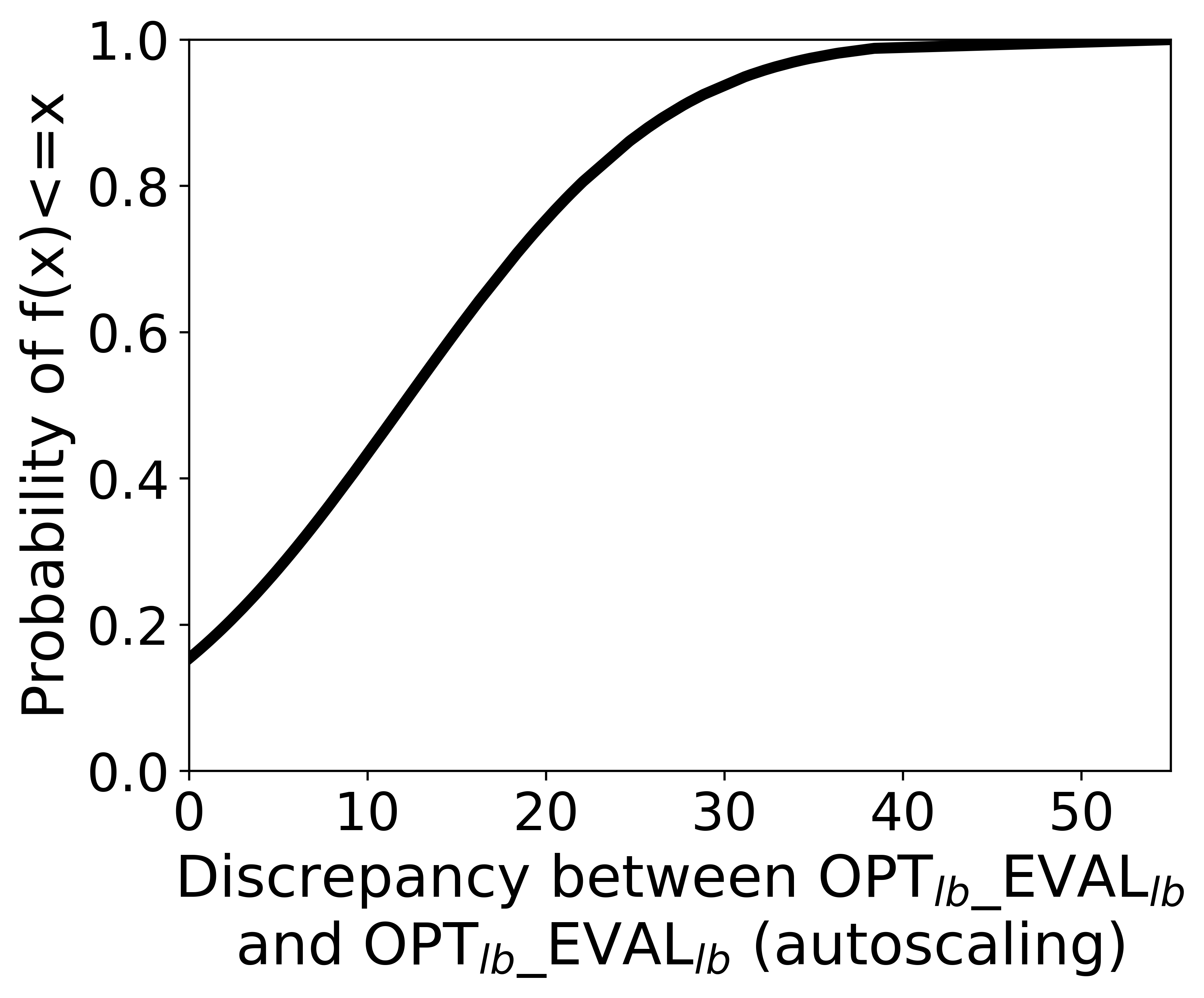}
    \end{center}
    \vspace{-0.2cm}
       \caption{\emph{CDF plot showing that the discrepancies due to carbon-aware resource autoscaling can be up to $55.1\%$ when analyzed over Electricity Maps'~\cite{emap} 2022 data.}}
    \label{fig:autoscale-discrepancy-cdf}
    \vspace{-0.5cm}
\end{figure}

\textcolor{black}{Similar to Section~\ref{sec:lb-temporal}, we then apply CarbonScaler to over 100 regions to get a comprehensive idea about the discrepancies between location- and market-based evaluations. Figure~\ref{fig:autoscale-discrepancy-cdf} shows the CDF plot of the discrepancies. We see an average of $11.9 \%$, with the maximum discrepancy being $55.1 \%$. Again, regions with no solar and wind show no discrepancy, while regions with higher differences between $CI_{res}$ and $CI_{lb}$ generally show more discrepancies.}

\vspace{-0.2cm}
\subsection{Key Takeaways}


\begin{enumerate}[leftmargin=*]
\item \textcolor{black}{Not accounting for PPAs while trying to optimize for carbon emission reductions raises discrepancies in the reported carbon savings. Our analysis shows possible overestimation of up to $55.1\%$ \textcolor{black}{for consumers without PPAs} when carbon optimizations and evaluations do not consider PPAs.}
\item \textcolor{black}{In some cases, this overestimation can also hide that not considering PPAs may possibly increase carbon emissions under the market-based method, giving consumers a false impression of carbon savings. For example, 3 out of 5 AWS regions analyzed in Section~\ref{sec:lb-autoscale} showed an increase in carbon emissions, while location-based evaluation still showed carbon savings.}
\item Since $CI_{res}$ differs from $CI_{lb}$ both spatially and temporally, the discrepancies vary across regions and with time. Even for regions and times with small discrepancies, the magnitude of added carbon emissions due to these overestimations may still possibly be significant, especially for large-scale consumers.
\item \textcolor{black}{The overestimation may be less for consumers whose demands are partially met by PPAs since the discrepancy will only be for the fraction of demand met using the residual grid mix. However, a detailed analysis is required to quantify the overestimation in such cases, which is left as future work. }
\end{enumerate}

\section{Carbon-savings with market-based attribution}
\label{sec:savings-ppa-aware}
Today, most grids have a portion of electricity contracted out via PPAs, and so the carbon savings will be less than reported via the location-based method. \textcolor{black}{In this section, we highlight what will happen if carbon reduction techniques both optimize and evaluate using market-based method (\mbmb) and calculate their carbon savings. \textcolor{black}{Again, we only show this from the viewpoint of consumers without any investments (i.e., $CI_{res} = CI_{mb}$).} In the extreme case of no PPAs, the optimization decisions and carbon savings will be exactly the same as \lblb since location-based and market-based methods will estimate the same carbon intensity if no PPAs exist. However, that is not the case in reality; some electricity is under PPA in most grids. Since we do not have information about the exact percentage of electricity under PPA in a region, we show our analysis over a range of percentages.}

\begin{figure}[t]
    \begin{center}
       \includegraphics[width=0.9\linewidth]{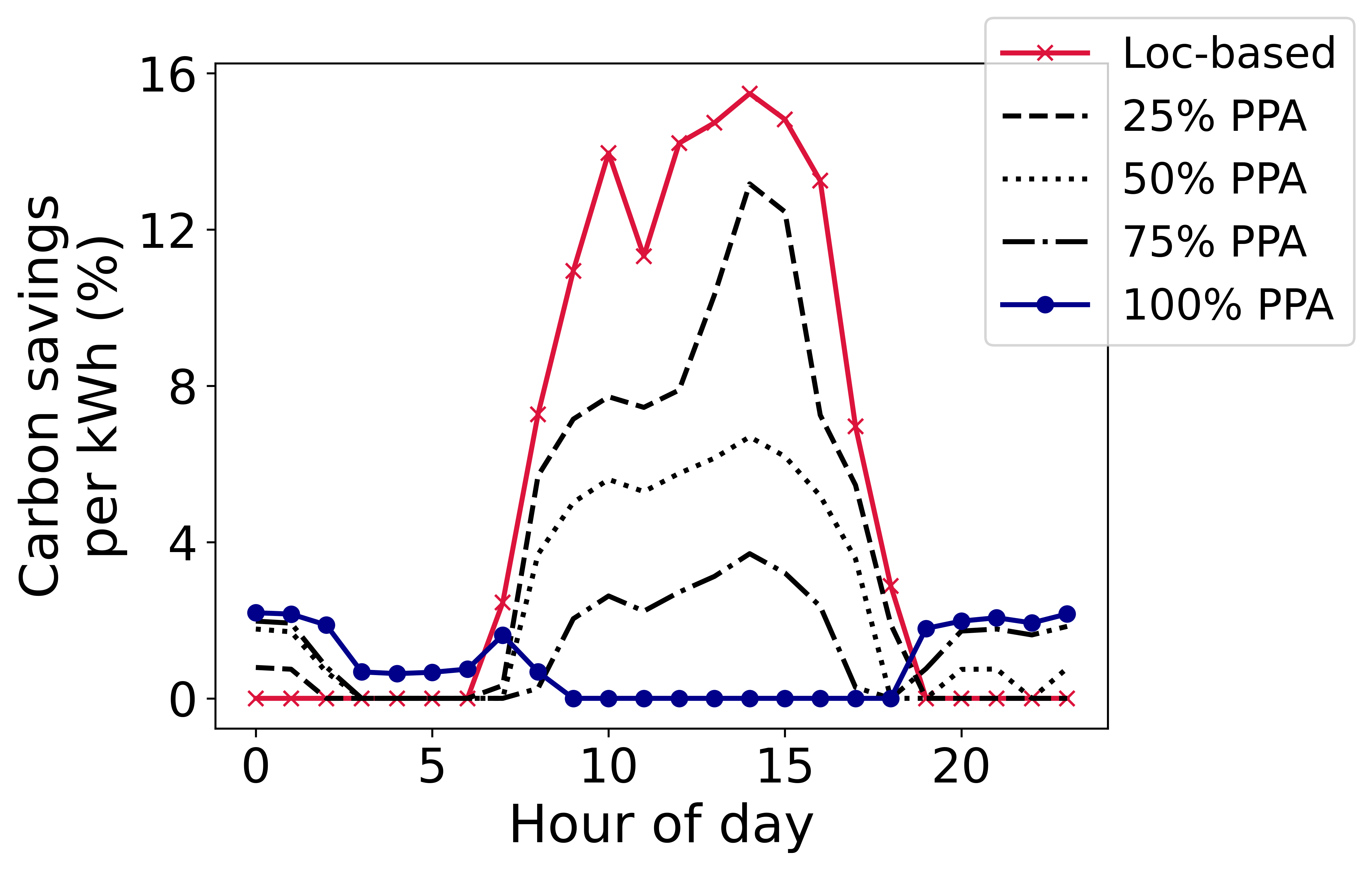}
    \end{center}
    \vspace{-0.2cm}
       \caption{\emph{Carbon savings are less for a consumer without PPA when using market-based attribution than when using location-based attribution. The savings also decrease with an increasing PPA \%.}}
    \label{fig:spatial-mb-carbon-savings}
    \vspace{-0.5cm}
\end{figure}

\begin{figure}[t]
    \begin{center}
       \includegraphics[width=0.9\linewidth]{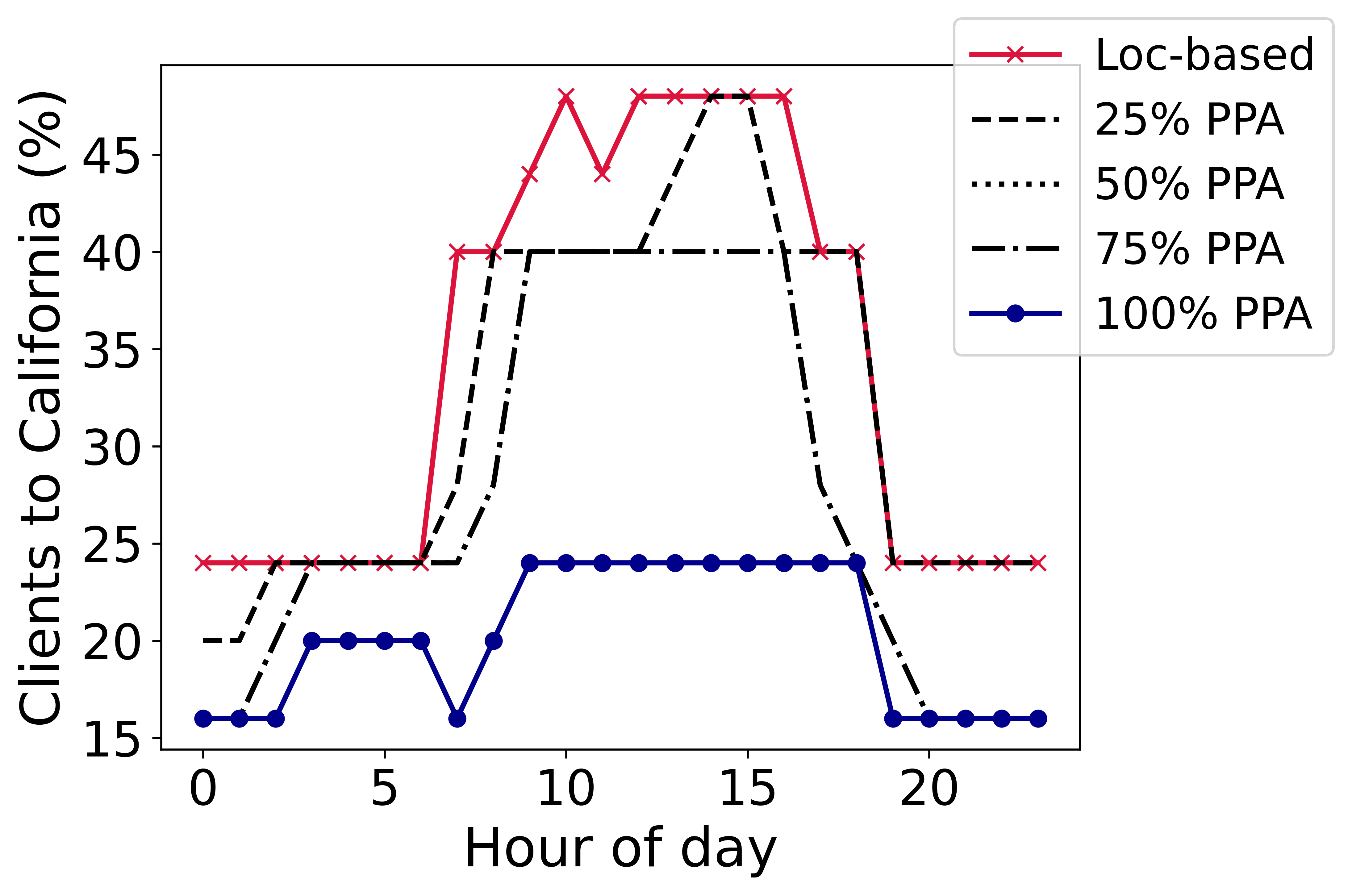}
    \end{center}
    \vspace{-0.2cm}
       \caption{\emph{Fewer clients are redirected to California when using the market-based method, as $CI_{res}$ increases with PPA \%.}}
    \label{fig:spatial-mb-client-redirect}
    \vspace{-0.5cm}
\end{figure}

\subsection{Spatial Load Shifting}
\label{sec:mb-spatial}
\textcolor{black}{Since $CI_{res}$ is always more than $CI_{lb}$, the carbon savings from spatially shifting workloads to greener regions under the market-based method are less than that obtained under the location-based method \textcolor{black}{for consumers without PPAs}. These savings reduce even further as the PPA percentage increases (refer Figure~\ref{fig:spatial-mb-carbon-savings}). While the location-based method (\lblb) can reduce carbon emissions by up to $15.4\%$, the maximum carbon savings is $6.7\%$ with $50\%$ PPA. If all renewables are under PPA, the maximum carbon reduction on the day is only $2.2\%$, which is $13.2 \%$ less carbon savings compared to \lblb.} In scenarios like these, moving client requests to greener regions may not be beneficial enough if such movement has high overheads.

We also observe that the carbon savings are obtained during different hours of the day when different attribution methods are considered. While the location-based optimization is more intuitive and shows savings during the day when solar energy is available, the market-based optimization achieves savings towards dawn and evening as more renewables are purchased. This is because the load balancer cannot redirect more clients to California during the day if solar is under PPA, and hence, must send those clients to other locations. Consequently, we see different decisions taken by the carbon-aware load balancer under the market-based method. For example, Figure~\ref{fig:spatial-mb-client-redirect} shows that at $100\%$ PPA, approximately $50\%$ fewer clients are redirected to California during the day when compared to the location-based method as in that case $CI_{res}$ of CAISO is much higher than $CI_{lb}$. Since the optimizer accounts for PPAs, there is no discrepancy during evaluation, and carbon emissions are never higher than the baseline, unlike in Figure~\ref{fig:spatial_best_worst}.

\begin{figure}[t]
    \begin{center}
       \includegraphics[width=0.85\linewidth]{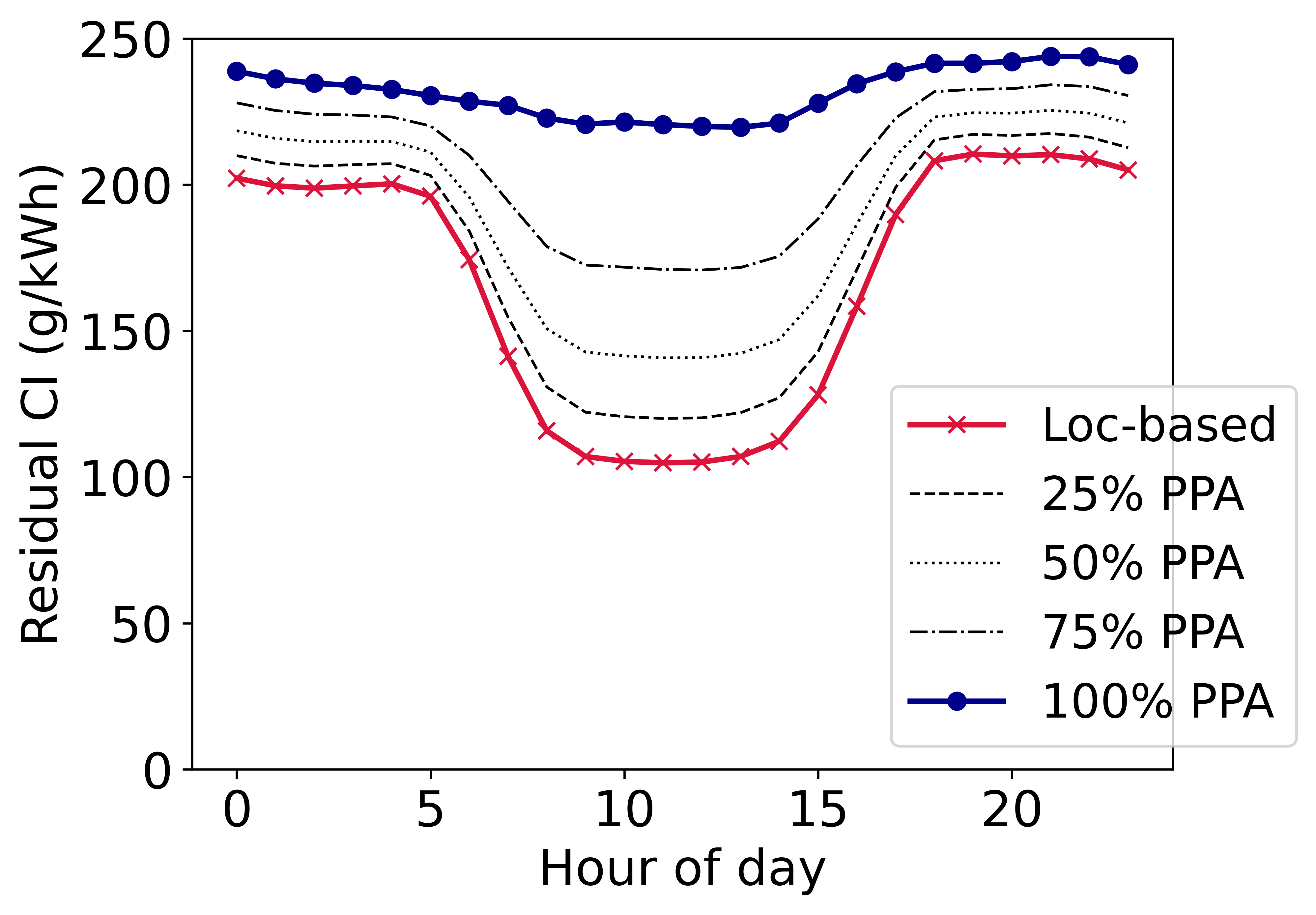}
    \end{center}
    \vspace{-0.2cm}
       \caption{\emph{Temporal variations in carbon intensity decrease as the residual grid electricity gets more brown.}}
    \label{fig:ciso_daily_variation}
    \vspace{-0.5cm}
\end{figure}

\subsection{Temporal Load Shifting}
\label{sec:mb-temporal}
\begin{figure}[t]
    \begin{center}
       \includegraphics[width=0.85\linewidth]{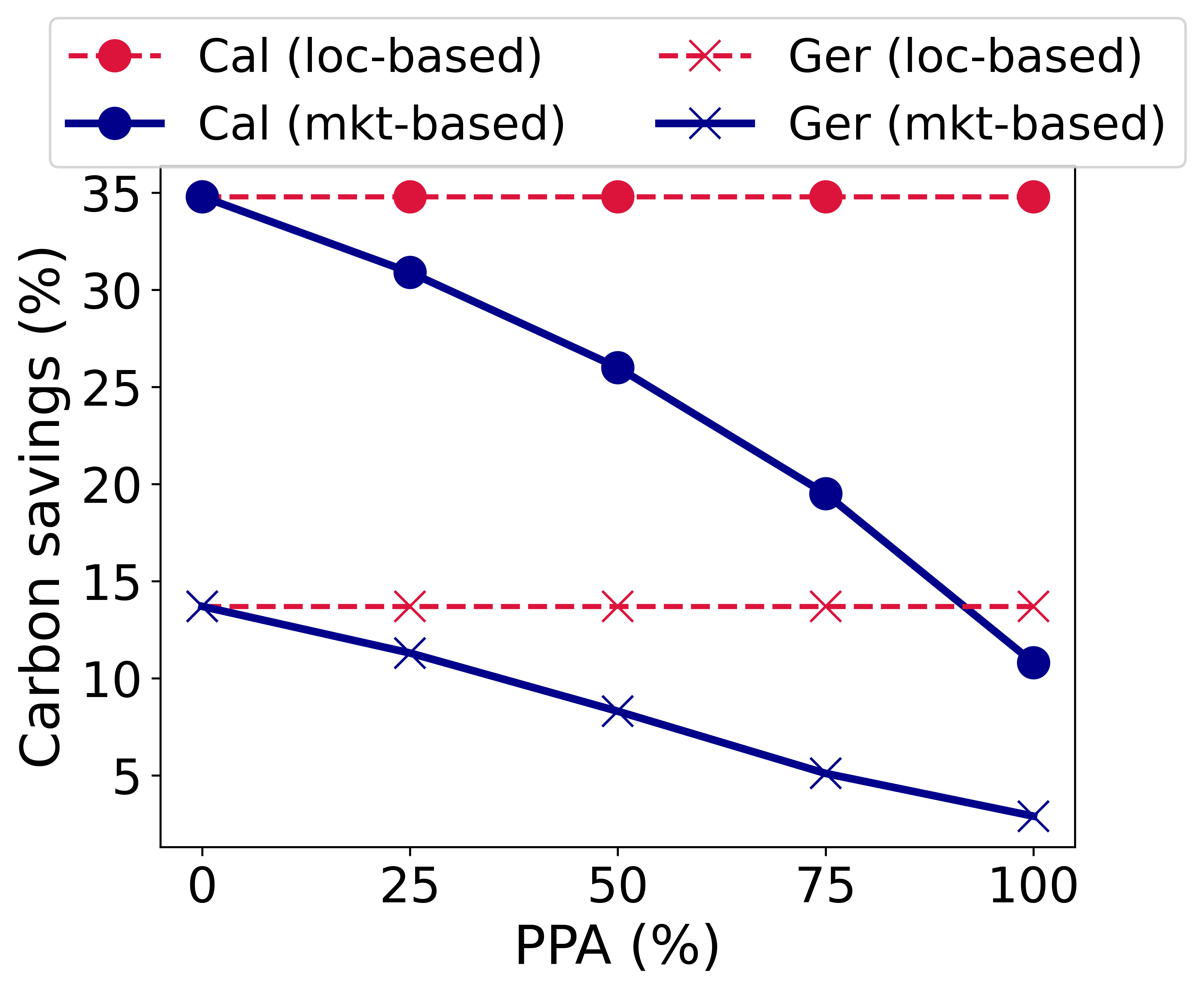}
    \end{center}
    \vspace{-0.2cm}
       \caption{\emph{Carbon savings due to temporal shifting of flexible jobs decrease with an increase in the PPA \%.}}
    \label{fig:temporal-mb-savings}
    \vspace{-0.3cm}
\end{figure}

\textcolor{black}{Temporal variations of carbon intensity decrease with an increasing PPA percentage since the source mix consists mostly of non-renewable sources, which are non-volatile. Figure~\ref{fig:ciso_daily_variation} shows this effect in California. Consequently, \textcolor{black}{for consumers without PPAs,} carbon savings due to the temporal shifting of flexible workloads decrease with an increasing PPA percentage under the market-based method, as shown in Figure~\ref{fig:temporal-mb-savings}. If all renewables are under PPA, the carbon-aware job scheduling algorithm used in Section~\ref{sec:lb-temporal} can achieve only $10.8\%$ (resp., $2.9\%$) with a $\pm 8$-hour flexibility window in California (resp., Germany). Thus, using market-based method (\mbmb) may potentially yield up to $24 \%$ (resp., $10.8 \%$) less carbon savings compared to the location-based method (\lblb).}

Note that in Germany, although the savings decrease, emissions are never more than a carbon-unaware scheduler, unlike in Section~\ref{sec:lb-temporal}. This is because the scheduler now both optimizes and evaluates using the market-based method (\mbmb). Thus, there are no discrepancies, and the consumers know the exact amount of emission reductions. 

Although there are no discrepancies, since the carbon savings in different regions may be negligible when most of the renewables are under PPA, it may not be beneficial to shift flexible workloads temporally if such shifting has a high overhead. In that case, either executing the job as soon as it arrives or other optimization techniques like spatial shifting may be more beneficial. 

\subsection{Resource Autoscaling}
\label{sec:mb-autoscale}
\textcolor{black}{Finally, we show the effects of PPA on carbon-aware autoscaling \textcolor{black}{for consumers without PPAs}. Like Sections~\ref{sec:mb-spatial} and \ref{sec:mb-temporal}, the carbon savings decrease as the PPA percentage increases. When all renewables are under PPA, market-based method (\mbmb) may yield up to $28.2 \%$ less carbon savings than the location-based method (\lblb) across the regions (refer Table~\ref{tab:scale-mb-savings}). The only exception is Ontario, where both location- and market-based method achieve similar savings as $CI_{res}$ is very close to $CI_{lb}$ due to very little solar and wind.} 

\begin{table}[t]
    \centering
    \begin{tabular}[h] { |>{\centering\arraybackslash}m{5em}||>{\centering\arraybackslash}m{4em}||>{\centering\arraybackslash}m{2em}|>{\centering\arraybackslash}m{2em}|>{\centering\arraybackslash}m{2em}|>{\centering\arraybackslash}m{2em}|}
     \hline
     \rule{0pt}{4ex} \textbf{Region} & {\textbf{Location-based}} & \multicolumn{4}{c|}{\textbf{Market-based}}\\
     \cline{2-6}
     \textbf{} & \textbf{} & \textbf{25\%} & \textbf{50\%} & \textbf{75\%} & \textbf{100\%} \\
     \hline
     \hline
     \textbf{Texas} & 11.3 & 9.2 & 6.8 & 3.9 & 0.8 \\
     \hline
     \textbf{New South Wales} & 18.2 & 14.2 & 9.6 & 5.2 & 3.5 \\
     \hline
     \textbf{Germany} & 24.7 & 20.7 & 15.6 & 9.4 & 1.0 \\
     \hline
     \textbf{California} & 35.7 & 31.0 & 25.2 & 16.8 & 7.5 \\
     \hline
     \textbf{Ontario} & 55.9 & 55.7 & 55.7 & 55.5 & 55.3 \\
     \hline
    \end{tabular}
    \caption{\emph{Carbon savings due to autoscaling decrease as residual grid electricity gets more brown with an increasing PPA \%.}}
    \label{tab:scale-mb-savings}
    \vspace{-0.5cm}
\end{table}

\begin{figure}[t]
    \begin{center}
       \includegraphics[width=\linewidth]{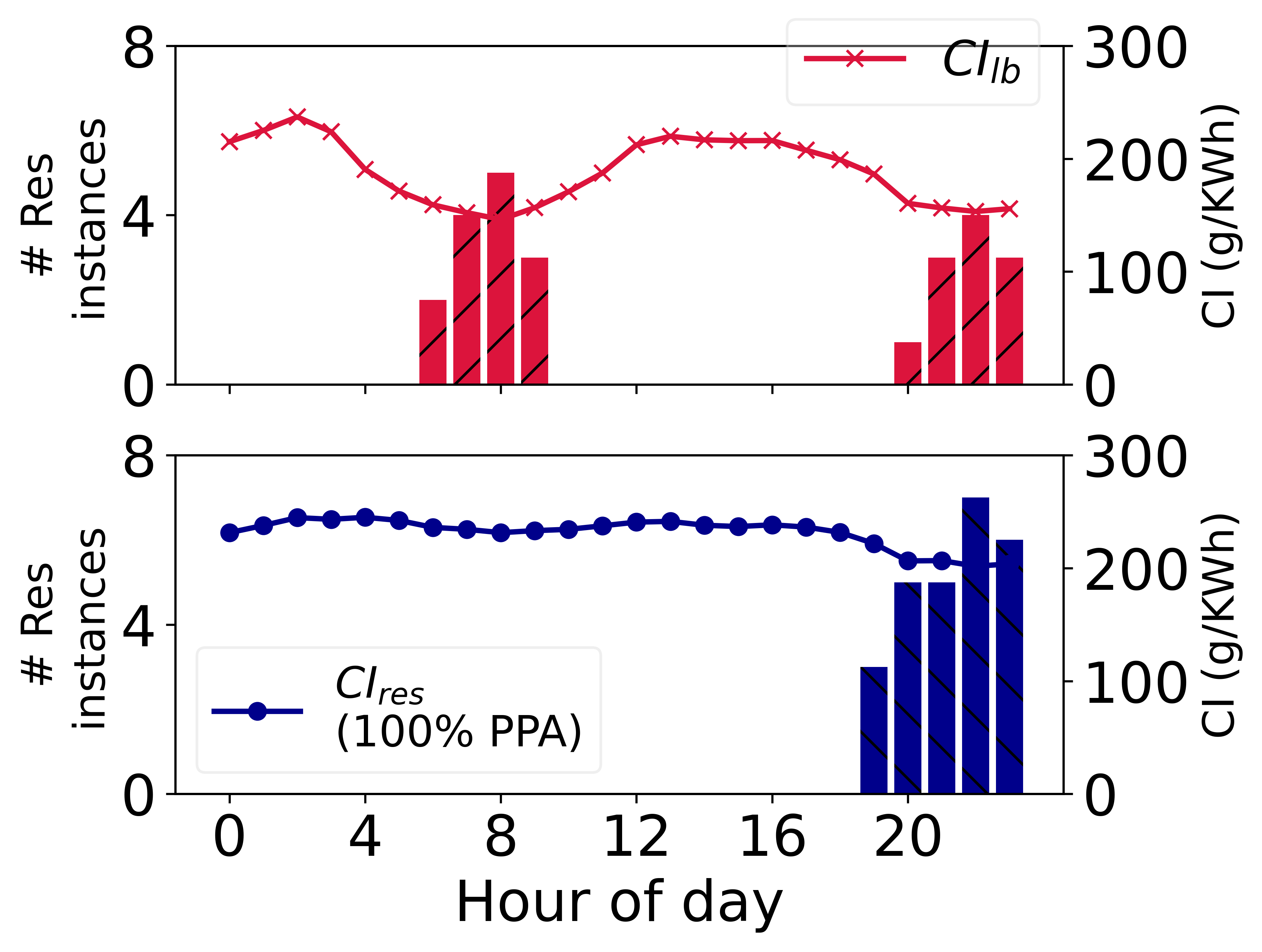}
    \end{center}
    \vspace{-0.2cm}
       \caption{\emph{CarbonScaler chooses different times to scale up resources when accounting for PPAs compared to the location-based method.}}
    \label{fig:scale-mb-schedule}
    \vspace{-0.3cm}
\end{figure}

Since CarbonScaler now follows $CI_{res}$ to allocate resources, resources are scaled at different times than the location-based method (\lblb), as shown in Figure~\ref{fig:scale-mb-schedule}. In the location-based method, CarbonScaler allocates more resources to cloud applications during hours 6--10 and 20--23 due to a dip in $CI_{lb}$. However, assuming all renewables are contracted out, $CI_{res}$ dips only during hours 19-23. Hence, when CarbonScaler accounts for PPAs while optimizing, it allocates more resources only during those hours.

Importantly, because of the above behaviour change, even though the carbon savings are reduced, CarbonScaler never increases emissions compared to a baseline that does not do autoscaling. For example, even with 100\% PPA, CarbonScaler achieves $3.5\%$ carbon savings in New South Wales (NSW), unlike in Section~\ref{sec:lb-autoscale}.

\subsection{Key Takeaways}

\begin{enumerate}[leftmargin=*]
\item For consumers without PPAs, carbon savings under the market-based method are usually lower than the location-based method. Our analysis shows that the market-based method may potentially yield up to $28.2 \%$ less carbon savings than the location-based method in the regions we have considered. \textcolor{black}{For consumers with PPAs, the savings may be higher since a portion of their demand is met using purchased low-carbon power, and that portion will have zero emissions. However, more analysis is required to quantify the savings in such cases.}

\item Carbon optimization decisions are different for the market-based method than the location-based method. The decisions using a market-based method may also not always be as intuitive as shifting demand to low-carbon regions or periods. Since $CI_{res}$ of a region can significantly vary from $CI_{lb}$, optimizers may opt to shift demand away from seemingly green regions which have high $CI_{res}$ due to PPAs. For example, in Section~\ref{sec:mb-spatial}, the carbon-aware load balancer redirected fewer clients to California during the day as California had high $CI_{res}$.

\item For consumers without PPAs, since the carbon savings may be negligible compared to carbon-unaware baselines, optimizing for carbon may not always be beneficial if the overhead of such optimization is high. For example, if there are minimal temporal variations, executing a job as soon as it arrives may be better than waiting for a low-carbon period.

\item The carbon optimization technique that holds for a location-based method may not be best suited when using the market-based method. Taking the above case of minimal temporal variations, shifting the job to a low-carbon region may be better. However, such analysis is kept as future work.
\end{enumerate}

\section{Related work}
\label{sec:related_work}
As the amount of PPAs and RECs increases, a few recent works highlight the need for accurate green energy attribution and carbon intensity calculation, and show the risks of inaccurate carbon accounting. Holzapfel et al.~\cite{holzapfel2023electricity} show how using location-based and market-based accounting in parallel can lead to discrepancies and overestimation of carbon savings, and offers potential solutions against that. Google~\cite{google_247_cfe} also mentions discrepancies due to a lack of residual electricity data when they account for their carbon emissions. Bjorn et al.~\cite{bjorn2022renewable} discuss how inaccurate use of RECs and PPAs provides a sense of inflated emission reduction estimates. While these works only mention discrepancies, we quantify that using current carbon optimization systems.

Brander et al.~\cite{brander2018creative} and Electricity Maps~\cite{emap_green_contracts} highlight the pitfalls of market-based accounting and recommend using the location-based accounting method. We do not recommend any specific method in our paper. Instead, we show the different amounts of carbon reductions obtained under the two methods.

There have also been numerous works on trying to reduce carbon emissions using carbon-aware spatial shifting, temporal shifting, or autoscaling~\cite{radovanovic2022carbon, wiesner2021let, maji2023bringing, gao2012s, hanafy2023carbonscaler, huber2021carbon, cheng2022carbon}. However, they only report emission reductions achieved under the location-based method. Our work shows that such reports can often overestimate carbon reductions due to the concurrent application of different attribution methods. We also show that when accounting for PPAs under the market-based method, the optimizers behave differently and the carbon reductions are much lower than reported.

\section{Discussions and Future Work}
\label{sec:discussions}
If everyone follows only location- or market-based method, there is no discrepancy~\cite{holzapfel2023electricity}. However, given the increasing trend of buying PPAs and the push towards 24/7 carbon-free energy~\cite{24_7_cfe, google_247_cfe}, the market-based method is more likely to be followed in practice.

\textcolor{black}{For the market-based method, information about the amount of electricity under PPA is crucial to calculate the residual grid mix and consequently estimate $CI_{res}$ and $CI_{mb}$. However, PPA agreements are usually neither visible to the electric grid nor easily accessible to the common public. Some commercial services track the amount of electricity under PPA~\cite{ceba_tracker, powerledger}, but such information is only available at a cost. Publicly accessible data about the residual grid mix is currently only available annually and lags for a few years~\cite{aib_residual, green_e_residual}. Thus, if we are to move towards decarbonization without any discrepancies, there is an urgent need for data about the amount of electricity under PPA in each regional electricity grid and the corresponding residual grid mix at an hourly scale. Recently, Electricity Maps~\cite{emap} and FlexiDAO~\cite{flexidao} have released a methodology for calculating hourly residual carbon intensity~\cite{hourly_res_ci}. While this is a step in the right direction, the data will still be available after a lag of 12--18 months~\cite{hourly_res_ci}. We need more efforts from both the buyers and sellers of PPAs to make such data available in real time and increase the visibility of the electrical grids and electricity markets to researchers and practitioners of carbon-aware computing.}

\textcolor{black}{If residual grid mix data is available in real time, the current consumption-based carbon intensity estimation algorithms~\cite{de2019tracking, tranberg2019real, horsch2018flow, tranberg2015power, bialek1996tracing, abdelkader2007transmission, kirschen1999tracing, schafer2019principal, li2013carbon} can be modified to incorporate PPAs and estimate $CI_{res}$ in real time. In that scenario, current online optimization algorithms will work without any modifications and report the true carbon reductions without any discrepancies, provided everyone follows $CI_{mb}$ to evaluate their respective carbon footprint. However, as mentioned, ensuring the availability of such data is not in our control. Thus, in the absence of such data, developing newer online algorithms that can work with partial carbon intensity information and are robust to uncertainties and inaccuracies in carbon intensity estimates is crucial. In the future, we also need algorithms that consider both location- and market-based carbon intensity --- that is, incorporate information about both the PPAs and the grid carbon intensity --- while making decisions, to reduce carbon emissions without any discrepancy.}

\section{Conclusions}
\label{sec:conclusion}
Many organizations have set ambitious targets to reduce carbon emissions as a part of their sustainability goals. Consumers rely on third-party services to modulate their electricity usage based on carbon intensity estimates and reduce their carbon footprint. Many consumers also use PPAs to offset their ``brown'' energy consumption. However, there are two carbon-free energy attribution methods in practice today and no consensus on which method to follow, resulting in a concurrent application of both estimates. 

In this paper, we show that such concurrent applications can cause discrepancies in the carbon savings reported by the current carbon reduction techniques. Our analysis across three state-of-the-art carbon reduction techniques shows a possible overestimation of up to $55.1\%$ in the carbon reductions reported by the consumers. In some cases, such overestimation can even hide increased emissions. We find that carbon reduction techniques behave differently under the market-based method than under the location-based method. Carbon reduction under the market-based method can also yield up to $28.2 \%$ less carbon reductions than those claimed using the location-based method for consumers without PPAs.

\begin{acks}
This work is supported in part by NSF grants 2105494, 2021693, 2213636 and 2020888, and a grant from VMware.
\end{acks}

\balance
\bibliographystyle{ACM-Reference-Format}
\bibliography{paper}


\end{document}